\newcommand{\bea}{\begin{eqnarray}}
\newcommand{\eea}{\end{eqnarray}}
\newcommand{\be}{\begin{equation}}
\newcommand{\ee}{\end{equation}}
\newcommand{\ben}{\begin{enumerate}}
\newcommand{\een}{\end{enumerate}}
\newcommand{\bi}{\begin{itemize}}
\newcommand{\ei}{\end{itemize}}
\newcommand{\bmi}[1]{\begin{minipage}{#1 cm}}
\newcommand{\emi}{\end{minipage}}

\def\eck#1{\left\lbrack #1 \right\rbrack}

\def\rund#1{\left( #1 \right)}
\def\abs#1{\left\vert #1 \right\vert}
\def\wave#1{\left\lbrace #1 \right\rbrace}
\def\ave#1{\left\langle #1 \right\rangle}

\def\Re{{\cal R}\hbox{e}}
\def\Im{{\cal I}\hbox{m}}

\def\P{{\cal P}}

\def\d{{\rm d}}

\def\eps{{\epsilon}}

\def\arcminf {\hbox{$.\!\!^{\prime}$}}
\def\vp{\varphi}
\def\vt{{\vartheta}}

\def\Real{{\rm I\mathchoice{\kern-0.70mm}{\kern-0.70mm}{\kern-0.65mm}%
  {\kern-0.50mm}R}}
\def\C{\rm C\kern-.42em\vrule width.03em height.58em depth-.02em
       \kern.4em}
\font \bolditalics = cmmib10
\def\bx#1{\leavevmode\thinspace\hbox{\vrule\vtop{\vbox{\hrule\kern1pt
        \hbox{\vphantom{\tt/}\thinspace{\bf#1}\thinspace}}
      \kern1pt\hrule}\vrule}\thinspace}

\def \vc #1{{\textfont1=\bolditalics \hbox{$\bf#1$}}}
{\catcode`\@=11
\gdef\SchlangeUnter#1#2{\lower2pt\vbox{\baselineskip 0pt \lineskip0pt
  \ialign{$\m@th#1\hfil##\hfil$\crcr#2\crcr\sim\crcr}}}
}
\def\gtrsim{\mathrel{\mathpalette\SchlangeUnter>}}
\def\lesssim{\mathrel{\mathpalette\SchlangeUnter<}}

\def\ueber#1#2{{\setbox0=\hbox{$#1$}%
  \setbox1=\hbox to\wd0{\hss$\scriptscriptstyle #2$\hss}%
  \offinterlineskip
  \vbox{\box1\kern0.4mm\box0}}{}}

\def\bx#1{\leavevmode\thinspace\hbox{\vrule\vtop{\vbox{\hrule\kern1pt
        \hbox{\vphantom{\tt/}\thinspace{\bf#1}\thinspace}}
      \kern1pt\hrule}\vrule}\thinspace}

\def\arcminf {\hbox{$.\!\!^{\prime}$}}                                     
\newcommand{\vectii}[2]{\rund{\begin{array}{c} #1 \\ #2 \end{array} }}

\def\elabel#1{\label{eq:#1}}

{\catcode`\@=11
\gdef\SchlangeUnter#1#2{\lower2pt\vbox{\baselineskip 0pt \lineskip0pt
  \ialign{$\m@th#1\hfil##\hfil$\crcr#2\crcr\sim\crcr}}}
}
\def\gtrsim{\mathrel{\mathpalette\SchlangeUnter>}}
\def\lesssim{\mathrel{\mathpalette\SchlangeUnter<}}

\documentclass[onecolumn]{aa}
\usepackage{graphicx}
\begin{document}
   \title{Analysis of two-point statistics of cosmic shear:
   I. Estimators and covariances}

   \author{Peter Schneider
          \inst{1,2}
          \and
          Ludovic van Waerbeke\inst{3,4}
	  \and
	  Martin Kilbinger\inst{1} 
	  \and
	  Yannick Mellier\inst{3,5}
          }

   \offprints{P. Schneider}

   \institute{Institut f. Astrophysik u. Extr. Forschung, Universit\"at Bonn,
              Auf dem H\"ugel 71, D-53121 Bonn, Germany\\
              \email{peter@astro.uni-bonn.de}
         \and
 	 	Max-Planck-Institut f. Astrophysik, Postfach 1317,
              D-85741 Garching, Germany
	\and
               Institute d'Astrophysique de Paris, 98 bis, boulevard
              Arago, F-75014 Paris, France
         \and
		Canadian Institute for Theoretical Astrophysics, 60 St
              Georges Str., Toronto, M5S 3H8 Ontario, Canada
	\and
		Observatoire de Paris, DEMIRM/LERMA, 61 avenue de 
               l'Observatoire, F-75014 Paris, France
             }

   \date{Received ; accepted }

   \abstract{Recently, cosmic shear, the weak lensing effect by the
   inhomogeneous matter distribution in the Universe, has not only
   been detected by several groups, but the observational results have
   been used to derive constraints on cosmological parameters. For
   this purpose, several cosmic shear statistics have been
   employed. As shown recently, all two-point statistical measures can
   be expressed in terms of the two-point correlation functions of the
   shear, which thus represents the basic quantity; also, from a
   practical point-of-view, the two-point correlation functions are
   easiest to obtain from observational data which typically have
   complicated geometry. We derive in this paper expressions for the
   covariance matrix of the cosmic shear two-point correlation
   functions which are readily applied to any survey
   geometry. Furthermore, we consider the more special case of a
   simple survey geometry which allows us to obtain approximations for
   the covariance matrix in terms of integrals which are readily
   evaluated numerically. These results are then used to study the
   covariance of the aperture mass dispersion which has been employed
   earlier in quantitative cosmic shear analyses. We show that the
   aperture mass dispersion, measured at two different angular scales,
   quickly decorrelates with the ratio of the scales. Inverting the
   relation between the shear two-point correlation functions and the
   power spectrum of the underlying projected matter distribution, we
   construct estimators for the power spectrum and for the band
   powers, and show that they yields accurate approximations; in
   particular, the correlation between band powers at different wave
   numbers is quite weak. The covariance matrix of the shear
   correlation function is then used to investigate the expected
   accuracy of cosmological parameter estimates from cosmic shear
   surveys. Depending on the use of prior information, e.g. from CMB
   measurements, cosmic shear can yield very accurate determinations
   of several cosmological parameters, in particular the normalization
   $\sigma_8$ of the power spectrum of the matter distribution, the
   matter density parameter $\Omega_{\rm m}$, and the shape parameter
   $\Gamma$.
	
   }

   \maketitle
%

\section{Introduction}
Cosmic shear, the distortion of the images of distant galaxies by the
tidal gravitational field of intervening matter inhomogeneities,
offers a direct way of probing the statistical properties of the
large-scale (dark) matter distribution in the Universe, without making
any assumption on the relation between dark and luminous matter (e.g.,
Blandford et al. 1991, Miralda-Escude 1991, Kaiser 1992, 1998, Jain \&
Seljak 1997, Bernardeau et al. 1997, Schneider et al. 1998, hereafter
SvWJK, van Waerbeke et al. 1999; Bartelmann \& Schneider 1999; Jain et
al. 2000, White \& Hu 2000; see Mellier 1999 and Bartelmann \&
Schneider 2001 for recent reviews). The first detections of cosmic
shear on wide-field imaging data (Bacon et al. 2000, 2002; Kaiser et
al. 2000; van Waerbeke et al. 2000, 2001, 2002; Wittman et al. 2000,
Maoli et al. 2001; Rhodes et al. 2001; H\"ammerle et al. 2002;
Hoekstra et al.\ 2002; Refregier et al.\ 2002) has demonstrated the
feasibility of this new window of observational cosmology, and yielded
already the first constraints on cosmological parameters, most
noticibly the normalization $\sigma_8$ of the dark matter power
spectrum, but also on the matter density parameter $\Omega_{\rm m}$
(van Waerbeke et al.\ 2002; Hoekstra et al.\ 2002).

Most analytical work on cosmic shear has been done on two-point
statistical measures of the distortion field, such as the shear
correlation functions, the shear variance in an apertures, or the
aperture mass (see Sect.\ts 2 for a definition of these
quantitities). Although higher-order statistical measures, such as the
skewness of the shear (Bernardeau et al.\ 1997), are likely to yield
additional, if not even superior constraints on cosmological
parameters, their theoretical predictions are more uncertain at
present. Recently, an estimator for the skewness of the shear was
developed (Bernardeau et al.\ 2002a), and applied to wide-field survey
data (Bernardeau et al\ 2002b), yielding a significant detection.

In this paper we consider second-order statistical measures only. All
of them can be derived in terms of the correlation functions, as shown
in, e.g., Crittenden et al.\ (2002, hereafter C02) and Schneider et
al.\ (2002, hereafter SvWM), and since the measurement of the
correlation functions is easier in practice than the other two-point
statistics (e.g., gaps in the data are easily dealt with), we consider
the correlation functions as the fundamental observables from a cosmic
shear survey. In order to use them for determining cosmological
parameters, it is important to know a practical and unbiased estimator
for them, and to determine the covariance of this estimator. Two
effects enter this covariance: a random part, which is due to the
intrinsic ellipticity of the galaxies from which the shear is
measured, together with measurement noise, and sampling (or cosmic)
variance. The first of these effects is expected to dominate on small
angular scales, whereas the second determines the covariance for large
separations. The covariance will depend not only on the total survey
area, but also on the survey geometry. As has been pointed out by
Kaiser (1998), in order to decrease the sampling variance on large
scales, it may be favourable to choose a survey geometry that samples
those scales sparsely. In order to design an optimized survey
geometry, the covariance as a function of survey geometry needs to be
calculated.

Here, we calculate the covariance matrices for the shear correlation
functions binned in angular separation. In Sect.\ts 2, we introduce
our notation and briefly summarize the two-point cosmic shear measures
and their interrelations. Unbiased estimators of the two basic
correlation functions are derived in Sect.\ts 3, togther with the
corresponding unbiased estimators of the aperture mass and the shear
dispersion. The covariance matrices of these correlation function
estimators are then derived in Sect.\ts 4, expressed in terms of a set
of galaxy positions. From these expressions, the covariances can be
determined for an arbitrary survey geometry. In a forthcoming paper
(Kilbinger et al., in preparation), we shall use the results of
Sect.\ts 4 to design an optimized geometry for a planned cosmic shear
survey. 

For the case of a filled survey geometry, the ensemble average of
these covariances can be further analyzed; using a few
approximations, we express in Sect.\ts 5 the covariances for this case
in terms of integrals. The corresponding expressions have been
evaluated, for a particular cosmological model, and are illustrated in
a set of figures. In Sect.\ts 6 we derive the covariance for the
aperture mass dispersion, which can be expressed simply in terms of the
covariances of the correlation functions. The variance of the aperture
mass dispersion, as well as the covariance, is then explicitly
calculated for a survey with filled geometry, showing that indeed the
aperture mass at two angular scales decorrelates quickly as the scale
ratio decreases away from unity. 

We then turn in Sect.\ts 7 to a simple estimator of the power spectrum
of the projected cosmic density field, which can be expresed in terms
of the correlation functions. Since the correlation functions will be
known only over a finite range in angular separation, the simple
estimator we derive is biased. We show that, provided the angular
range on which the correlation functions can be measured is as large
as can be expected with the next generation of cosmic shear surveys,
this bias is indeed {\it very} small over a large range of wave
numbers. We derive the covariance of the power spectrum estimator and
calculate it explicitly for the filled survey geometry case; the
resulting error bars on the estimated power spectrum are quite a bit
smaller than one might have expected, given the simplicity of the
approach. In Sect.\ts 8 we consider the accuracy with which the
parameters of the cosmological model can be constrained, given a
survey area. In fact, by fitting the correlation function directly to
model predictions, even the currently available cosmic shear surveys
can yield fairly accurate constraints on cosmological parameter. 
Finally, we summarize our results in Sect.\ 9.

In this paper we shall assume that the observable shear is due to the
tidal gravitational field of the cosmological matter distribution
only; in this case, the two shear components are not mutually
independent. This is due to the fact that the gravitational field is a
gradient field. In the language of some recent papers (e.g., C02; Pen
et al.\ 2002; SvWM), we thus assume that the shear field is a pure
E-mode field. B-modes, (or the `curl component'), can in principle be
generated if the intrinsic orientation of the galaxies from which the
shear is measured are correlated, e.g. due to tidal interactions of
dark matter halos in which these galaxies are formed (Croft \& Metzler
2000; Pen et al. 2000; Heavens et al. 2001; Catelan et al. 2001;
Mackey et al. 2002; Brown et al. 2002). Also, the clustering of source
galaxies in redshift space generates a B-mode contribution which,
however, turns out to be fairly small (SvWM).  This restriction to
E-modes only affects the interrelations between various two-point
statistics. Inclusion of B-modes would not change the results of
Sects.\ 3 through 5, and much of Sects.\ 6 and 7 will also be left
unaffected in the presence of a B-mode contribution; we shall indicate
this in due course.

\section{Two-point measures of cosmic shear}
We follow here the notation of Bartelmann \& Schneider (2001). The
power spectrum of the projected density field is denoted by
$P_\kappa(\ell)$, where $\ell$ is the Fourier variable on the sky. 

The shear correlation functions are defined by considering pairs of
positions $\vc\vt$ and $\vc\theta+\vc\vt$, and defining the
tangential and cross-component of the shear $\gamma=\gamma_1+{\rm
i}\gamma_2$ at position $\vc\vt$ for this pair as
\be
\gamma_{\rm t}=-\Re\rund{\gamma\,{\rm e}^{-2{\rm i}\vp}} \qquad ;\qquad
\gamma_{\times}=-\Im\rund{\gamma\,{\rm e}^{-2{\rm i}\vp}} \;,
\elabel{1}
\ee
where $\vp$ is the polar angle of the separation vector $\vc
\theta$. Then we define the two shear correlation functions (e.g.,
Kaiser 1992)
\be
\xi_\pm(\theta):=
\ave{\gamma_{\rm t}\gamma_{\rm t}} \pm\ave{\gamma_\times \gamma_\times}
={1\over 2\pi}\int_0^\infty\d \ell\;\ell\, P_\kappa(\ell)\,{\rm
J}_{0,4}(\ell\theta)\;,
\elabel{2}
\ee
where the Bessel function ${\rm J}_0$ (${\rm J}_4$) corresponds to the
`+' (`$-$') correlation function.  The shear dispersion in a circle of
radius $\theta$ is defined by considering circular
apertures and ensemble-averaging over the square of the complex shear;
its relation to the power spectrum reads (e.g., Kaiser 1992) 
\be
\ave{\abs{\gamma}^2}(\theta)={1\over 2 \pi}
\int_0^\infty{\d \ell\; \ell}\,P_\kappa(\ell) {4 \eck{{\rm
J}_1(\ell\theta)}^2\over (\ell\theta)^2} \;.
\elabel{3}
\ee
Furthermore, the aperture mass $M_{\rm ap}$ in an aperture of radius
$\theta$ is defined as a weighted average over the tangential shear
component (see, e.g. Schneider 1996; SvwJK), and its dispersion is
related to the power spectrum by
\be
\ave{M_{\rm ap}^2}(\theta)
={1\over 2\pi}\int_0^\infty{\d \ell\; \ell}\,P_\kappa(\ell)
{576 \eck{{\rm J}_4(\ell\theta)}^2\over (\ell\theta)^4} \;. 
\elabel{4}
\ee
where the same weight function in the definition of $M_{\rm ap}$ as in
SvWM was assumed.

All these 2-point statistics are thus linearly filtered versions of
the power spectrum $P_\kappa$, where the filter functions are quite
different between the various statistics. For the correlation
function $\xi_+$, the filter function ${\rm J}_0(\eta)$ is very broad,
about constant for $\eta\ll 1$, and oscillating for large $\eta$, with
an amplitude decreasing as $\eta^{-1/2}$. The filter function ${\rm
J}_4(\eta)$ for $\xi_-$ has the same slow decrease, but behaves as
$\eta^4$ for small $\eta$, and is therefore more localized than the one
for $\xi_+$. The filter function for the shear dispersion, $[2 {\rm
J}_1(\eta)/\eta]^2$, is a low-pass filter, i.e. constant for $\eta\ll 1$, and
then decreasing in amplitude as $\eta^{-3}$ for large $\eta$. Finally,
the filter function for the aperture mass dispersion is $[24 {\rm
J}_4(\eta)/\eta^2]^2$ and thus behaves like $\eta^4$ for small $\eta$,
and decreases oscillatory as $\eta^{-5}$ for $\eta\to \infty$. Hence, 
$\ave{M_{\rm ap}^2}$ yields the most local estimate of the underlying
power spectrum of the projected mass. On the other hand, because it is
so localized, it contains less power in its filter, so that the value
of  $\ave{M_{\rm ap}^2}$ is smaller than that of
$\ave{\abs{\gamma}^2}$ on the same filter scale $\theta$. 

The various two-point statistics of the shear are related to each
other; in particular, they can all be expressed in terms of the
correlation functions, as was shown in C02, Pen
et al.\ (2002) and
SvWM. We briefly summarize the results here.

Making use of the orthogonality of Bessel functions,
the power spectrum can be expressed in terms of the correlation
functions $\xi_+(\theta)$ and $\xi_-(\theta)$, by multiplying
eqs.(\ref{eq:2}) by $\theta\,{\rm J}_0(\ell\theta)$ and
$\theta\,{\rm J}_4(\ell\theta)$, respectively, and then integrating over
$\theta$, to obtain 
\be
P_\kappa(\ell)=2\pi\int_0^\infty\d\theta\;\theta\,\xi_+(\theta)\,{\rm
J}_0(\ell\theta) =2\pi\int_0^\infty\d\theta\;\theta\,\xi_-(\theta)\,{\rm
J}_4(\ell\theta) \;.
\elabel{7}
\ee
These equations express the power spectrum directly in terms of
the observable correlation function; however, in order to evaluate
$P_\kappa(\ell)$ from them, one would need to know the correlation
functions for all angles. In Sect.\ts 7 below, we shall investigate
how well the power spectrum can be determined from knowing the
correlation function over a limited range of separations.

The two equations (\ref{eq:2}) and (\ref{eq:7}) allow us to express
$\xi_+$ in terms of $\xi_-$, and reversely (see SvWM for a derivation),
\be
\xi_-(\theta)=\xi_+(\theta)+\int_0^\theta\d\vt\;\vt\,
\xi_+(\vt)\rund{{4\over\theta^2}-{12\vt^2\over
\theta^4}}\qquad ;\qquad 
\xi_+(\theta)=\xi_-(\theta)+\int_\theta^\infty\d\vt\;\vt\,\xi_-(\vt)
\rund{{4\over\vt^2}-{12\theta^2\over
\vt^4}}\;.
\elabel{N4}
\ee
Hence one can obtain $\xi_-(\theta)$ from the correlation function
$\xi_+(\vt)$ in the interval $0\le \vt\le \theta$, and so this
relation can be directly applied to observational data, with a minor
extrapolation to small separations. The reverse
relation, expressing $\xi_+$ in terms of $\xi_-$, is less useful in
practice, owing to the infinite range of integration. 

Next we express the shear dispersion (\ref{eq:3}) in terms of the
correlation function, by inserting (\ref{eq:7}) into
(\ref{eq:3}); this yields (van Waerbeke 2000; SvWM)
\be
\ave{\abs{\gamma}^2}(\theta)=\int_0^{2\theta} {\d\vp\;\vp\over
\theta^2}\, \xi_+(\vp)\,S_+\rund{\vp\over \theta}
=\int_0^\infty {\d\vp\;\vp\over
\theta^2}\, \xi_-(\vp)\,S_-\rund{\vp\over \theta}\;,
\label{eq:S2}
\ee
where
\bea
S_+(x)&=&4\int_0^\infty{\d t\over t}\,{\rm J}_0(x t)\,\eck{{\rm
J}_1(t)}^2 = {1\over\pi}\eck{4\arccos\rund{x\over
2}-x\sqrt{4-x^2}}\,{\rm H}(2-x)
\;, \nonumber \\
S_-(x)&=&4\int_0^\infty{\d t\over t}\,{\rm J}_4(x t)\,\eck{{\rm
J}_1(t)}^2 ={x\sqrt{4-x^2}(6-x^2)-8(3-x^2)\arcsin(x/2)
\over \pi x^4} {\rm H}(2-x)+{4(x^2-3)\over x^4}\,{\rm H}(x-2)\;,
\eea
and ${\rm H}(x)$ is the Heaviside step function.
Hence, the function $S_+(x)$ vanishes for $x>2$, so that the shear
dispersion can be expressed as a finite-range integral over the
correlation function $\xi_+$; this is not the case for $S_-$.

Similarly, one can express the aperture mass dispersion in terms of
the correlation functions, by inserting (\ref{eq:7}) into
(\ref{eq:4}):
\be
\ave{M_{\rm ap}^2}(\theta)=\int_0^{2\theta} {\d\vp\;\vp\over
\theta^2}\, \xi_+(\vp)\,T_+\rund{\vp\over \theta}
=\int_0^{2\theta} {\d\vp\;\vp\over
\theta^2}\, \xi_-(\vp)\, T_-\rund{\vp\over\theta}\;,
\elabel{T2}
\ee
where
\bea
T_+(x)&=&576\int_0^\infty{\d t\over t^3}\,{\rm J}_0(x t)\,\eck{{\rm
J}_4(t)}^2  \nonumber \\
&=&\wave{{6(2-15x^2)\over 5}\eck{1-{2\over\pi}\arcsin\rund{x\over 2}}
+ {x\sqrt{4-x^2}\over 100\pi}
\rund{120+2320x^2-754x^4+132 x^6-9x^8}}{\rm H}(2-x)
\;,\\
T_-(x)&=&576\int_0^\infty{\d t\over t^3}\,{\rm J}_4(x t)\,\eck{{\rm
J}_4(t)}^2 ={192\over 35\pi}x^3\rund{1-{x^2\over 4}}^{7/2}\,{\rm H}(2-x)
\;. \nonumber
\eea
Both of these functions vanish for $x>2$, so that $\ave{M_{\rm ap}^2}$
can be expressed by a finite integral over either $\xi_\pm$.
The functions $S_\pm$ and $T_\pm$ are plotted in SvWM.

The fact that we can express the shear dispersion and the aperture
mass dispersion directly in terms of the correlation function over a
finite interval is expected, given that the
estimators of both statistics include products of ellipticities of
pairs of objects separated by no more than the diameter of the
aperture. However, what could not have been guessed a priori is that
$\ave{M_{\rm ap}^2}(\theta)$ can be expressed by a finite integral
over $\xi_+$ and $\xi_-$ separately.
The determination of these statistics in terms of the
correlation function is in practice easier than laying down apertures,
owing to the holes and gaps in a data set; in addition, a comparison
of the directly determined shear and aperture mass dispersion with
those obtained from the correlation functions yields a useful
check on the integrity of the data.

\section{Estimators}
We shall now consider practical estimators of the correlation functions
and the other two-point statistics.
The observable ellipticity $\eps_i$ of a galaxy image at angular
position $\vc\theta_i$ is related to the intrinsic ellipticity
$\eps^{\rm s}_i$ and the shear $\gamma(\vc\theta_i)$ by
\be
\eps_i=\eps^{\rm s}_i+\gamma(\vc\theta_i)\;,
\label{eq:N10}
\ee
where it has been assumed that $\abs{\gamma}\ll 1$ for this weak
lensing relation to be valid. In addition to an ellipticity, each
galaxy may carry a weight factor $w_i$ which reflects the precision
with which its ellipticity can be determined -- noisy objects can then
be downweighted by assigning small values of $w_i$ to them (Hoekstra
et al.\ 2000; Erben et al.\ 2001; Bacon et al.\ 2001; Pen et al.\
2002).  We shall assume that the correlation function is to be
estimated in bins of angular width $\Delta\vt$, and define the
function $\Delta_\vt(\phi)=1$ for
$\vt-\Delta\vt/2<\phi\le\vt+\Delta\vt/2$, and zero otherwise; hence,
$\Delta_\vt(\phi)$ defines the bin at angle $\vt$.  An estimator for
the correlation function $\xi_+(\vt)$ is then
\be
\hat \xi_+(\vt)={ \sum_{ij} w_i\,w_j\,(\eps_{i{\rm t}}\eps_{j{\rm t}}
+ \eps_{i\times}\eps_{j\times})\,
\Delta_\vt(\abs{\vc\theta_i-\vc\theta_j})
\over
N_{\rm p}(\vt)}\;,
\qquad
N_{\rm p}(\vt)=\sum_{ij}
w_i\,w_j\,\Delta_\vt(\abs{\vc\theta_i-\vc\theta_j}) \;;
\elabel{N12}
\ee
$N_{\rm p}(\vt)$ is the effective `number of pairs' in the bin
considered (in fact, if all weights are unity, $N_{\rm p}$ is twice
the number of pairs), and the tangential and cross components of the
ellipticity are defined in analogy to the corresponding shear
components in (\ref{eq:1}).  The expectation value of this estimator
is obtained by averaging over the source ellipticities, assumed to be
randomly oriented, and an ensemble average of the shear field. Since
\be
\ave{\eps_{i{\rm t}}\eps_{j{\rm t}}
+ \eps_{i\times}\eps_{j\times}}=\sigma_\eps^2\delta_{ij}
+\xi_+(\abs{\vc\theta_i-\vc\theta_j})\;, 
\label{eq:N13}
\ee
where $\sigma_\eps^2$ is the dispersion of the intrinsic galaxy
ellipticity, we see immediately that $\hat \xi_+$ is an unbiased
estimator of $\xi_+$,
\be
\ave{\hat \xi_+(\vt)}= \xi_+(\vt)\;,
\label{eq:N14}
\ee
since the product
$\delta_{ij}\Delta_\vt(\abs{\vc\theta_i-\vc\theta_j})$ vanishes for
all pairs $i,j$. Analogously, an unbiased estimator for $\xi_-$ is 
\be
\hat \xi_-={ \sum_{ij} w_i\,w_j\,\rund{\eps_{i{\rm t}}\eps_{j{\rm t}}
- \eps_{i\times}\eps_{j\times}} \,
\Delta_\vt(\abs{\vc\theta_i-\vc\theta_j})
\over
N_{\rm p}(\vt)}\;,\qquad \ave{\hat \xi_-(\vt)}= \xi_-(\vt)\;.
\ee

Next we obtain an unbiased estimator for the aperture mass
dispersion. For that we assume that the centers of the bins on
which the correlation function is calculated are described by
$\vt_i=(i-1/2)\Delta\vt$, and that the aperture radius $\theta$ is an
integer multiple of the bin width, $\theta=m\,\Delta\vt$. Then, the
integrals in (\ref{eq:T2}) are replaced by sums over the bins,
yielding an estimator for $\ave{M_{\rm ap}^2(\theta)}$,
\be
{\mathcal M}(\theta)
={\Delta\vt\over \theta^2}\sum_{i=1}^{2m}\vt_i\,\eck{K_+ \hat \xi_+(\vt_i)
T_+\rund{\vt_i\over\theta} + (1-K_+) \hat \xi_-(\vt_i)
T_-\rund{\vt_i\over\theta} }\;,
\elabel{N23}
\ee
where $K_+$ describes the relative contributions of the two
expressions (\ref{eq:T2}). For example, in the presence of B-modes,
one needs to use $K_+=1/2$ (SvWM). Similarily, an unbiased estimator
for the shear dispersion is
\be
{\mathcal S}(\theta)={\Delta\vt\over \theta^2}
\eck{K_+ \sum_{i=1}^{2m}\vt_i\, \hat \xi_+(\vt_i)
S_+\rund{\vt_i\over\theta} +
(1-K_+)\sum_{i=1}^\infty \vt_i\,\hat \xi_-(\vt_i)
S_-\rund{\vt_i\over\theta} }\;;
\ee
again, in the presence of B-modes, $K_+=1/2$ shall be chosen (however,
the infinite support of $S_-$ requires the knowledge of the
correlation function $\xi_-$ for all separations). Both of
the above are unbiased estimators, $\ave{{\mathcal
M}(\theta)}=\ave{M_{\rm ap}^2(\theta)}$, $\ave{{\mathcal S}(\theta)
}=\ave{\abs{\gamma}^2}(\theta)$.

\section{Covariance of the estimators}

Next we calculate the covariance of the various estimators, starting
with the correlation functions. Hence, we define 
\be
{\rm Cov}(\hat\xi_\pm,\vt_1;\hat\xi_\pm,\vt_2)
:=\ave{\rund{\hat \xi_\pm(\vt_1)-\xi_\pm(\vt_1)}\rund{\hat
\xi_\pm(\vt_2)-\xi_\pm(\vt_2)}} \;.
\ee
Consider first the `++'-covariance function, for which one needs to
evaluate 
\be
\ave{\hat \xi_+(\vt_1)\hat \xi_+(\vt_2)}
={1\over N_{\rm p}(\vt_1)N_{\rm p}(\vt_2)}
{\sum_{ijkl}w_iw_jw_kw_l
\Delta_{\vt_1}(ij) \Delta_{\vt_2}(kl)
\ave{\rund{\eps_{i1}\eps_{j1}+\eps_{i2}\eps_{j2}} 
 \rund{\eps_{k1}\eps_{l1}+\eps_{k2}\eps_{l2}}}}   \;,
\elabel{N15}
\ee
where we defined
$\Delta_\vt(ij)\equiv\Delta_{\vt}(\abs{\vc\theta_i-\vc\theta_j})$ and
used the fact that $\eps_{i{\rm t}}\eps_{j{\rm t}} +
\eps_{i\times}\eps_{j\times}=\eps_{i1}\eps_{j1}+\eps_{i2}\eps_{j2}$.
Next, the four-point correlation function of the ellipticities needs
to be evaluated. For that, we use (\ref{eq:N10}) and expand the
resulting expression. Only terms of even power in $\eps^{\rm s}$ and
$\gamma$ survive the averaging over the source ellipticities and the
ensemble average. Then,
\bea
\ave{\eps_{i\alpha}\eps_{j\beta}\eps_{k\mu}\eps_{l\nu}}
&=&\ave{\gamma_{i\alpha}\gamma_{j\beta}\gamma_{k\mu}\gamma_{l\nu}}
+{\sigma_\eps^2\over 2}\Bigl(
\delta_{jl}\delta_{\beta\nu}\ave{\gamma_{i\alpha}\gamma_{k\mu}}
+\delta_{jk}\delta_{\beta\mu}\ave{\gamma_{i\alpha}\gamma_{l\nu}}
+\delta_{il}\delta_{\alpha\nu}\ave{\gamma_{j\beta}\gamma_{k\mu}}
+\delta_{ik}\delta_{\alpha\mu}\ave{\gamma_{j\beta}\gamma_{l\nu}} \Bigr)
\nonumber \\
&+&\ave{\eps^{\rm s}_{i\alpha}\eps^{\rm s}_{j\beta}
\eps^{\rm s}_{k\mu}\eps^{\rm s}_{l\nu}}\;,
\eea
valid for $i\ne j$ and $k\ne l$, as needed in (\ref{eq:N15}); here,
Greek letters are $\in\{1,2\}$. To evaluate the four-point function of
the shear, we shall assume that the shear field is Gaussian, so that
the four-point function can be written as a sum over products of
two-point functions. We shall later comment on the effect this
assuption has on the determination of the covariances. The four-point
function of the intrinsic ellipticity also factorizes, since at most
two of the indices $i,j,k,l$ are equal. Therefore,
\bea
\ave{\eps_{i\alpha}\eps_{j\beta}\eps_{k\mu}\eps_{l\nu}}
={\sigma_\eps^2\over 2} \Bigl(
\delta_{jl}\delta_{\beta\nu}\ave{\gamma_{i\alpha}\gamma_{k\mu}}
+\delta_{jk}\delta_{\beta\mu}\ave{\gamma_{i\alpha}\gamma_{l\nu}}
+\delta_{il}\delta_{\alpha\nu}\ave{\gamma_{j\beta}\gamma_{k\mu}}
\!\!&+&\!\!\delta_{ik}\delta_{\alpha\mu}\ave{\gamma_{j\beta}\gamma_{l\nu}}
\Bigr) 
\nonumber \\
+
\ave{\gamma_{i\alpha}\gamma_{j\beta}}\ave{\gamma_{k\mu}\gamma_{l\nu}}
+\ave{\gamma_{i\alpha}\gamma_{k\mu}}\ave{\gamma_{j\beta}\gamma_{l\nu}}
+\ave{\gamma_{i\alpha}\gamma_{l\nu}}\ave{\gamma_{j\beta}\gamma_{k\mu}}
\!\!&+&\!\!\rund{\sigma_\eps^2\over 2}^2
\rund{\delta_{ik}\delta_{jl}\delta_{\alpha\mu}\delta_{\beta\nu}
+\delta_{il}\delta_{jk}\delta_{\alpha\nu}\delta_{\beta\mu}}\;.
\eea 
The correlation functions of the shear components can be expressed as
(Kaiser 1992)
\be
\ave{\gamma_{i1}\gamma_{j1}}={1\over2}\eck{\xi_+(ij)+\xi_-(ij)\cos(4\vp_{ij})}
\; ;\;
\ave{\gamma_{i2}\gamma_{j2}}={1\over2}\eck{\xi_+(ij)-\xi_-(ij)\cos(4\vp_{ij})}
\; ;\;
\ave{\gamma_{i1}\gamma_{j2}}={1\over2}\xi_-(ij)\sin(4\vp_{ij})\;,
\ee
where we have written
$\xi_\pm(ij)\equiv\xi_\pm(|\vc\theta_i-\vc\theta_j|)$, and $\vp_{ij}$
is the polar angle of the difference vector
$\vc\theta_i-\vc\theta_j$. From these relations, one obtains for the
covariance matrix
\bea
{\rm Cov}(\hat\xi_+,\vt_1;\hat\xi_+,\vt_2)&=&
{1\over N_{\rm p}(\vt_1)N_{\rm p}(\vt_2)}
\Biggl[ \sigma_\eps^4\bar\delta(\vt_1-\vt_2)
\sum_{ij}w_i^2w_j^2\Delta_{\vt_1}(ij)
+2\sigma_\eps^2\sum_{ijk}w_i^2w_jw_k \Delta_{\vt_1}(ij)
\Delta_{\vt_2}(ik) \xi_+(jk) \nonumber \\
&+&\sum_{ijkl}w_iw_jw_kw_l
\Delta_{\vt_1}(ij) \Delta_{\vt_2}(kl)
\Bigl(  \xi_+(il)\xi_+(jk)+\cos\eck{4\rund{\vp_{il}-\vp_{jk}}}
\xi_-(il)\xi_-(jk)\Bigr)\Biggr]\;,
\elabel{Cpp}
\eea
where the function $\bar\delta(\vt_1-\vt_2)$ is zero if the two
separation bins are different, and is 1 if they are the same. The
first term in (\ref{eq:Cpp}) therefore contributes only to the
diagonal terms in the covariance matrix. In the absence of shear
correlations, the covariance matrix would be diagonal; correlation
populates the off-diagonal elements of the covariance tensor.

To calculate the covariance matrix for the $\xi_-$ correlation
function, we first write 
\[
\eps_{i{\rm t}}\eps_{j{\rm t}} - \eps_{i\times}\eps_{j\times} 
=\rund{\eps_{i1}\eps_{j1}-\eps_{i2}\eps_{j2}}\cos 4\vp_{ij}
+\rund{\eps_{i1}\eps_{j2}+\eps_{i2}\eps_{j1}}\sin4\vp_{ij}\;;
\]
inserting this into the definition of the covariance matrix and
performing the same step as for the `++' covariance, one finds
\bea
{\rm Cov}(\hat\xi_-,\vt_1;\hat\xi_-,\vt_2)&=&
{1\over N_{\rm p}(\vt_1)N_{\rm p}(\vt_2)}
\Biggl[ \sigma_\eps^4\bar\delta(\vt_1-\vt_2)
\sum_{ij}w_i^2w_j^2\Delta_{\vt_1}(ij)\nonumber \\
&+&2\sigma_\eps^2\sum_{ijk}w_i^2w_jw_k \Delta_{\vt_1}(ij)
\Delta_{\vt_2}(ik) \cos\eck{4(\vp_{ij}-\vp_{ik})}\xi_+(jk)
\elabel{Cmm}\\ 
+\sum_{ijkl}w_iw_jw_kw_l     \!\!\!\!\! &&\!\!\!\!   
\Delta_{\vt_1}(ij) \Delta_{\vt_2}(kl)
\Bigl(\cos\eck{4(\vp_{ij}-\vp_{il}-\vp_{jk}+\vp_{kl})}
\xi_-(il)\xi_-(jk)+\cos\eck{4\rund{\vp_{ij}-\vp_{kl}}}
\xi_+(il)\xi_+(jk)\Bigr)\Biggr]\;.\nonumber
\eea
Finally, the mixed covariance matrix can be calculated in the same
manner, yielding
\bea
{\rm Cov}(\hat\xi_+,\vt_1;\hat\xi_-,\vt_2)&=&
{1\over N_{\rm p}(\vt_1)N_{\rm p}(\vt_2)}
\Biggl[ 2\sigma_\eps^2\sum_{ijk}w_i^2w_jw_k \Delta_{\vt_1}(ij)
\Delta_{\vt_2}(ik) \cos\eck{4(\vp_{ik}-\vp_{jk})}\xi_-(jk)\nonumber\\ 
&+&2\sum_{ijkl}w_iw_jw_kw_l  
\Delta_{\vt_1}(ij) \Delta_{\vt_2}(kl)
\cos\eck{4(\vp_{il}-\vp_{kl})}
\xi_-(il)\xi_+(jk)\Biggr]\;.
\elabel{Cpm}
\eea

\section{Averaging over an ensemble of galaxy positions}
Given a model for the shear correlation, the covariances
(\ref{eq:Cpp}--\ref{eq:Cpm}) can be calculated using the actual galaxy
positions and their weight factors; in principle, this procedure is
straightforward to apply to a given data set. Alternatively, given the
geometry of a data field, then by randomly distributing galaxy
positions the expected covariance of the shear correlation can be
determined; this procedure can be used to design and optimize cosmic
shear surveys (Kilbinger et al., in preparation). These calculations
are, however, time-consuming, given the sum over three and four galaxy
positions. It is therefore of interest to consider a relatively simple
situation where fairly explicit expressions for the covariance
matrices can be obtained. In particular, we shall calculate the
ensemble average of the covariance matrices.  We consider here a
survey geometry which consists of a single data field of solid angle
$A$ and galaxy number density $n$, so that $N=nA$ is the total number
of galaxies in the survey. The survey geometry is assumed to be
`simple', i.e.\ consisting of a simply connected region, say, a
quadratic field. We shall assume that all weight factors are unity,
$w_i=1$. Furthermore, we shall consider separations $\vt_i$ which are
small compared to the `diameter' of the survey field, $\vt_i^2\ll A$;
in this case, we can neglect `boundary effects' which otherwise would
complicate the analysis tremendously. With these assumptions, the
number of pairs in the bin characterized by $\vt$ is then
\be
N_{\rm p}(\vt)=A\,n\,2\pi\vt\,\Delta\vt\,n\;.
\elabel{N20}
\ee
The ensemble average over galaxy positions is carried out by the
averaging operator
\[
{\rm E}=\prod_{i=1}^N\rund{{1\over A}\int_A \d^2\theta_i}\;,
\]
i.e. the galaxies are assumed to be randomly placed on the field.
The average of the first term in (\ref{eq:Cpp}) and (\ref{eq:Cmm}) is
simple and reads
\be
{\rm E}\rund{{\sigma_\eps^4\bar\delta(\vt_1-\vt_2)
\over N_{\rm p}(\vt_1)N_{\rm p}(\vt_2)}
\sum_{ij}\Delta_{\vt_1}(ij)}
=
{\sigma_\eps^4\over N_{\rm p}(\vt_1)}\bar\delta(\vt_1-\vt_2)
=: D\,\bar\delta(\vt_1-\vt_2) \;.
\elabel{Ddef}
\ee
As expected, this term depends explicitly on the bin size chosen,
since $N_{\rm p}\propto \Delta\vt$. In practical units,
\[
D=3.979\times 10^{-9}\rund{\sigma_\eps\over 0.3}^4\rund{A\over 1\,{\rm
deg}^2}^{-1} \rund{n\over 30\,{\rm
arcmin}^{-2}}^{-2}\rund{\vt\over 1\,{\rm
arcmin}}^{-2}\rund{\Delta\vt/\vt\over 0.1}^{-1}\;.
\]
As we shall see, the other terms
are independent of the choice of the bins. To evaluate the other
terms, one notes that the expectation value of all terms involving a
sum over three galaxy positions can be written in terms of
\be
E_{abc}\equiv {\rm
E}\rund{\sum_{ijk}\Delta_{\vt_1}(ij)\,\Delta_{\vt_2}(ik)\, F_a(\vp_{ij})
\,F_b(\vp_{ik})\, F_c(\vc\theta_k-\vc\theta_j)}\;,
\elabel{abc}
\ee
or linear combinations thereof. Applying the averaging operator, we
note that there are $N(N-1)(N-2)\approx N^3$ permutations of galaxies,
so that 
\bea
E_{abc}&=&{N^3\over A^3}\int\d^2\theta_1\int\d^2\theta_2\int\d^2\theta_3
\;\Delta_{\vt_1}(12)\,\Delta_{\vt_2}(13)\, F_a(\vp_{12})
\,F_b(\vp_{13})\, F_c(\vc\theta_3-\vc\theta_2)\nonumber\\
&=&{N^3\over A^2}\int_{\vt_1-\Delta\vt/2}^{\vt_1+\Delta\vt/2}
\d\phi_1\,\phi_1\int_0^{2\pi}\d\vp_1\,F_a(\vp_1)
\int_{\vt_2-\Delta\vt/2}^{\vt_2+\Delta\vt/2}
\d\phi_2\,\phi_2\int_0^{2\pi}\d\vp_2\,F_b(\vp_2)
\;F_c(\vc\phi_2-\vc\phi_1) \;, \nonumber
\eea
where in the second step we have written
$\vc\theta_2=\vc\theta_1+\vc\phi_1$,
$\vc\theta_3=\vc\theta_1+\vc\phi_2$, 
so that $\vc\theta_3-\vc\theta_2=\vc\phi_2-\vc\phi_1$, $\vp_1$ and
$\vp_2$ denote the polar angles of $\vc\phi_1$, $\vc\phi_2$,
respectively. After this substitution, the integral becomes
independent of $\vc\theta_1$, which can be integrated to yield a
factor $A$. Assuming that the bin width $\Delta\vt$ is small, one can
evaluate the $\phi_1$ and $\phi_2$ integrals, to obtain 
\be
{E_{abc}\over N_{\rm p}(\vt_1) N_{\rm p}(\vt_2)}
={1\over (2\pi)^2 A n}
\int_0^{2\pi}\d\vp_1\,F_a(\vp_1)\int_0^{2\pi}\d\vp_2\,F_b(\vp_2)
\,F_c\vectii{\vt_2\cos\vp_2-\vt_1\cos\vp_1}
{\vt_2\sin\vp_2-\vt_1\sin\vp_1} \;. 
\elabel{Eabc}
\ee
Similarily, the expectation value of all terms involving a
sum over four galaxy positions can be written in terms of
\bea
E_{abcd}&\equiv& {\rm
E}\rund{\sum_{ijkl}\Delta_{\vt_1}(ij)\,\Delta_{\vt_2}(kl)\,
F_a(\vc\theta_k-\vc\theta_j)\,F_b(\vc\theta_l-\vc\theta_i)\,
F_c(\vp_{ij})\,F_d(\vp_{kl})}\nonumber \\
&=&{N^4\over A^4}
\int\d^2\theta_1\int\d^2\theta_2\,\Delta_{\vt_1}(12)\,F_c(\vp_{12})
\int\d^2\theta_3\int\d^2\theta_4\,\Delta_{\vt_2}(34)\,F_d(\vp_{34})\,
F_a(\vc\theta_3-\vc\theta_2)\,F_b(\vc\theta_4-\vc\theta_1) 
\nonumber \\
&=&{N^4\over A^3}\int\d^2\phi
\int_{\vt_1-\Delta\vt/2}^{\vt_1+\Delta\vt/2}
\d\phi_1\,\phi_1\int_0^{2\pi}\d\vp_1\,F_c(\vp_1)
\int_{\vt_2-\Delta\vt/2}^{\vt_2+\Delta\vt/2}
\d\phi_2\,\phi_2\int_0^{2\pi}\d\vp_2\,F_d(\vp_2)\,
F_a(\vc\phi-\vc\phi_1)\,
F_b(\vc\phi+\vc\phi_2)\;, \nonumber
\eea
where in the second step we defined
$\vc\theta_2=\vc\theta_1+\vc\phi_1$,
$\vc\theta_4=\vc\theta_3+\vc\phi_2$; then, the arguments of the
functions $F_a$ and $F_b$ become
$\vc\theta_3-\vc\theta_2=\vc\theta_3-\vc\theta_1-\vc\phi_1$ and
$\vc\theta_4-\vc\theta_1=\vc\theta_3-\vc\theta_1+\vc\phi_2$, i.e. they
depend only on the difference $\vc\phi= \vc\theta_3-\vc\theta_1$, so
that the integral over $\vc\theta_3$ can be carried out, yielding a
factor $A$. Performing the $\phi_1$ and $\phi_2$ integration, assuming
small bin width, one obtains
\be
{E_{abcd}\over N_{\rm p}(\vt_1) N_{\rm p}(\vt_2)}
={1\over (2\pi)^2 A }\int_0^\infty\d\phi\,\phi
\int_0^{2\pi}\d\vp_1\,F_c(\vp_1)\int_0^{2\pi}\d\vp_2\,F_d(\vp_2)\,
\int_0^{2\pi}\d\vp\,F_a(\vc\psi_a)\,F_b(\vc\psi_b)\;,
\elabel{Eabcd}
\ee
where the vectors
\be
\vc\psi_a=\vectii{\phi\cos\vp-\vt_1\cos\vp_1}{\phi\sin\vp-\vt_1\sin\vp_1}
\quad ;\quad
\vc\psi_b=\vectii{\phi\cos\vp+\vt_2\cos\vp_2}{\phi\sin\vp+\vt_2\sin\vp_2}
\elabel{thepsis}
\ee
have been defined for later convenience.

Next we shall evaluate the second term of (\ref{eq:Cpp}) which is of
the form (\ref{eq:abc}), with $F_a=1=F_b$,
$F_c(\vc\psi)=\xi_+(|\vc\psi|)$; inserting these expressions into
(\ref{eq:Eabc}), the expectation value of the second term in
(\ref{eq:Cpp}) becomes
\be
q_{++}={2\sigma_\eps^2\over \pi A n}
\int_0^\pi\d\vp\;\xi_+\rund{\sqrt{\vt_1^2+\vt_2^2-2\vt_1\vt_2\cos\vp}
}\;.
\elabel{qpp}
\ee
The expectation value of the third term in (\ref{eq:Cpp}) can be
calculated by splitting it up into three parts: in the first, 
$F_a(\vc\psi)=\xi_+(|\vc\psi|)=F_b(\vc\psi)$, and the other two are
obtained by expanding the cosine-term, so that for one of them,
$F_a(\vc\psi)=\cos 4\vp_\psi\,\xi_-(|\vc\psi|)=F_b(\vc\psi)$, and for
the other, cos is replaced by sin; here, $\vp_\psi$ is the polar angle
of $\vc\psi$. Note that all three terms have 
$F_c=1=F_d$. The final result then reads
\be
{\rm E}\rund{{\rm Cov(\hat\xi_+,\vt_1;\hat\xi_+,\vt_2)}}
=D\,\bar\delta(\vt_1-\vt_2)+q_{++}+r_{+0}+r_{+1}\;,
\ee
where $D$ and $q_{++}$ have been defined in (\ref{eq:Ddef}) and
(\ref{eq:qpp}), respectively, 
\bea
r_{+0}&=&{2\over \pi A}\int_0^\infty\d\phi\,\phi
\int_0^\pi\,\d\vp_1\,\xi_+(|\vc\psi_a|)
\int_0^\pi\,\d\vp_2\,\xi_+(|\vc\psi_b|) \;,
\nonumber \\
r_{+1}&=&{1\over (2\pi) A}\int_0^\infty\d\phi\,\phi
\int_0^{2\pi}\d\vp_1\,\xi_-(|\vc\psi_a|)
\int_0^{2\pi}\d\vp_2\,\xi_-(|\vc\psi_b|)
\eck{\cos 4\vp_a \,\cos 4\vp_b + \sin 4\vp_a \,\sin 4\vp_b}\;,
\eea
and $\vp_a$, $\vp_b$ are the polar angles of $\vc\psi_a$, $\vc\psi_b$,
respectively, $\cos 4\vp_a=1-8\psi_{a1}^2\psi_{a2}^2/|\vc\psi_a|^4$,
$\sin 4\vp_a=4\psi_{a1}\psi_{a2}(\psi_{a1}^2
-\psi_{a2}^2)/|\vc\psi_a|^4$, and the analogous expressions for
$\vp_b$. Note that the $\vp$-integration present in (\ref{eq:Eabcd})
has dropped out as the integrand depends only on $\vp_1-\vp$, and
$\vp_2-\vp$; hence, the $\vp$-integration can be carried out and one
can use $\vp=0$ in (\ref{eq:thepsis}).

Several issues are worth mentioning: (1) only the first term
containing the `delta function' depends on the bin width $\Delta\vt$;
thus, the bin width only affects the autovariance. (2) All terms are
proportional to $A^{-1}$; hence, the relative contribution of the
terms is independent of the survey area, at least for separations well
below the `diameter' of the survey area for which the foregoing
procedure of the ensemble averaging is valid. (3) The terms denoted by
`$r$' are independent of the intrinsic ellipticity dispersion and of
the number density of galaxies. Hence, these terms describe the cosmic
variance and thus provide a limit on the accuracy of the determination
of the correlation function for a given survey geometry, independent
of the observing conditions which determine $n$.

The expectation values of the other covariance matrices are calculated
in a similar manner. Consider the `$-$$-$' covariance (\ref{eq:Cmm})
next; the first term agrees with that of (\ref{eq:Cpp}). For the
second term, we can apply (\ref{eq:Eabc}), after expanding the cosine;
then $F_a(\vp)$ and $F_b(\vp)$ are either $\cos 4\vp$ or $\sin 4\vp$,
and $F_c(\vc\psi)=\xi_+(|\vc\psi|)$. Similarily, (\ref{eq:Eabcd})
can be applied to the third term of (\ref{eq:Cmm}), after expanding
the cosine; using (\ref{eq:Eabcd}) term by term, and combining the
results afterwards, one obtains
\be
{\rm E}\rund{{\rm Cov(\hat\xi_-,\vt_1;\hat\xi_-,\vt_2)}}
=D\,\bar\delta(\vt_1-\vt_2)+q_{--}+r_{-0}+r_{-1}\;,
\ee
where 
\bea
q_{--}&=&{2\sigma_\eps^2\over \pi A n}
\int_0^\pi\d\vp\;\xi_+\rund{\sqrt{\vt_1^2+\vt_2^2-2\vt_1\vt_2\cos\vp}
}\,\cos(4\vp)\;, \nonumber \\
r_{-0}&=&{1\over (2\pi) A}\int_0^\infty\d\phi\,\phi
\int_0^{2\pi}\d\vp_1\,\xi_-(|\vc\psi_a|)
\int_0^{2\pi}\d\vp_2\,\xi_-(|\vc\psi_b|)\;
\cos[4(\vp_1+\vp_2-\vp_a-\vp_b)]\;, \\
r_{-1}&=&{1\over (2\pi) A}\int_0^\infty\d\phi\,\phi
\int_0^{2\pi}\d\vp_1\,\xi_+(|\vc\psi_a|)
\int_0^{2\pi}\d\vp_2\,\xi_+(|\vc\psi_b|)\;
\cos[4(\vp_1-\vp_2)]\,.\nonumber
\eea
Finally, the expectation value of (\ref{eq:Cpm}), the mixed
covariance, is calculated in a similar manner, yielding
\be
{\rm E}\rund{{\rm Cov(\hat\xi_+,\vt_1;\hat\xi_-,\vt_2)}}
=q_{+-} + r_{+-}\;,
\ee
with
\bea
q_{+-}&=&{2\sigma_\eps^2\over \pi A n}
\int_0^\pi\d\vp\;\eck{\sum_{k=0}^4\vectii{4}{k} (-1)^k\,\vt_1^k\,
\vt_2^{4-k}\,\cos(k\vp)}\,\rund{\vt_1^2+\vt_2^2-2\vt_1\vt_2\cos\vp}^{-2}
\xi_-\rund{\sqrt{\vt_1^2+\vt_2^2-2\vt_1\vt_2\cos\vp}}\nonumber \\
r_{+-}&=&{1\over \pi A}\int_0^\infty\d\phi\,\phi
\int_0^{2\pi}\d\vp_1\,\xi_+(|\vc\psi_a|)
\int_0^{2\pi}\d\vp_2\,\xi_-(|\vc\psi_b|)\;
\cos[4(\vp_2-\vp_b)]\;.
\eea
Hence, as already seen from (\ref{eq:Cpm}), the cross-covariance
matrix has no pure noise term from the intrinsic galaxy ellipticity
dispersion as the other two covariance matrices. 

   \begin{figure}
   \centering
   \includegraphics[width=19cm]{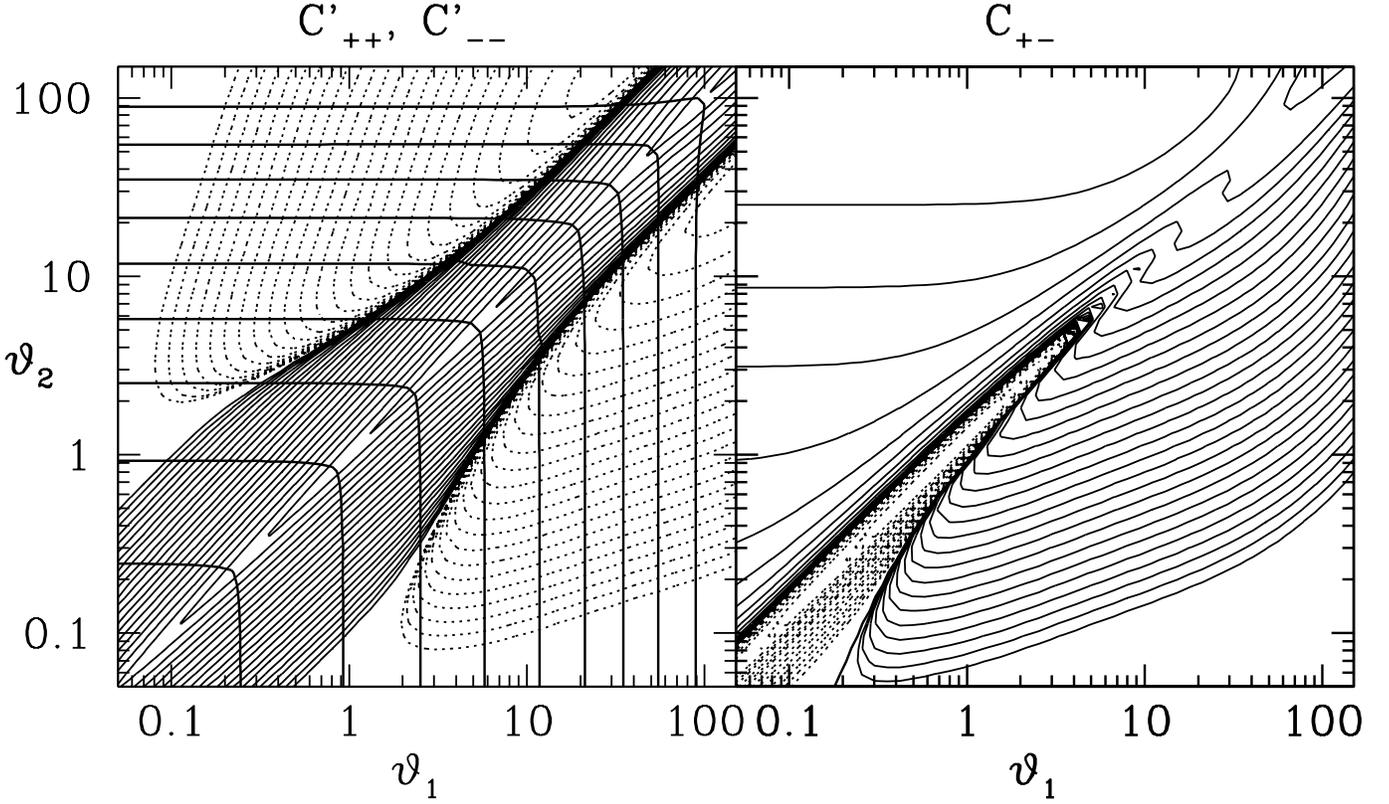}
      \caption{The correlation functions. In the left panel, we have
   plotted the covariance matrices ${\rm Cov}_{++}'(\vt_1,\vt_2)$
   (thick solid curves) and ${\rm Cov}_{--}'(\vt_1,\vt_2)$, i.e. the
   covariance matrices with the shot-noise term removed. For ${\rm
   Cov}_{++}'(\vt_1,\vt_2)$, the contours are linearly spaced, with
   the lowest value at $10^{-9}$ (outer-most contour) and highest
   value $9\times 10^{-9}$ for small $\vt_1$, $\vt_2$. For ${\rm
   Cov}_{--}'(\vt_1,\vt_2)$, contours are logarithmically spaced,
   with consecutive contours differing by a factor 1.5. The
   solid contours display positive values of ${\rm
   Cov}_{--}'(\vt_1,\vt_2)$, starting from $10^{-14}$, with the
   maximum value of $\sim 3\times 10^{-9}$ in the upper right corner,
   and dotted contours show negative values of ${\rm
   Cov}_{--}'(\vt_1,\vt_2)$, starting at $-10^{-15}$. In the right
   panel, ${\rm Cov}_{+-}(\vt_1,\vt_2)$ is shown, again with
   logarithmically spaced contours differing by a factor of 1.5. Solid
   contours are for positive values of ${\rm Cov}_{+-}(\vt_1,\vt_2)$,
   starting at $10^{-14}$, negative values are shown by dotted contours,
   starting at $-10^{-13}$. 
}
         \label{fig2}
   \end{figure}

We have obtained numerical estimates for the ensemble-averaged
covariance matrices derived above (see Fig.\ts\ref{fig2}). In the
numerical estimates given in this paper (except Sect.\ts 8.2), we have
used a standard set of parameters. The cosmological parameters are
those of a by-now standard $\Lambda$-dominated universe, $\Omega_{\rm
m}=0.3$, $\Omega_\Lambda=0.7$. The power spectrum of the density
fluctuations is described by its primordial slope of $n=1$, a shape
parameter $\Gamma=0.21$, and a normalization of $\sigma_8=1$. We used
the fit formula of Bardeen et al.\ (1986) for the linear power
spectrum, and the prescription of Hamilton et al.\ (1991) in the form
given in Peacock \& Dodds (1996) to describe the non-linear evolution
of the power spectrum. Furthermore, we fix the survey properties to be
described by a fiducial area of $A=1\,{\rm deg}^2$, a number density
$n=30\,{\rm arcmin}^{-2}$ of source galaxies, and an intrinsic
ellipticity dispersion of $\sigma_\eps=0.3$. The source galaxies were
assumed to have a redshift distribution $p(z)\propto
z^2\,\exp[-(z/z_0)^{1.5}]$, so that the mean redshift is $\bar z =
1.505\,z_0$. For the examples shown in Sects.\ts 5 through 7, we take
$z_0=1$. Note that all covariances simply scale with $A^{-1}$, so that
the results displayed here are easily translated to other survey
sizes. This scaling is also implicitly implied when considering scales
of order a degree or more -- all the numerical estimates are for the
ensemble averaged covariances, and their validity as given here depends
on the assumption $\theta^2\ll A$.

   \begin{figure}
   \centering
   \includegraphics[width=17cm]{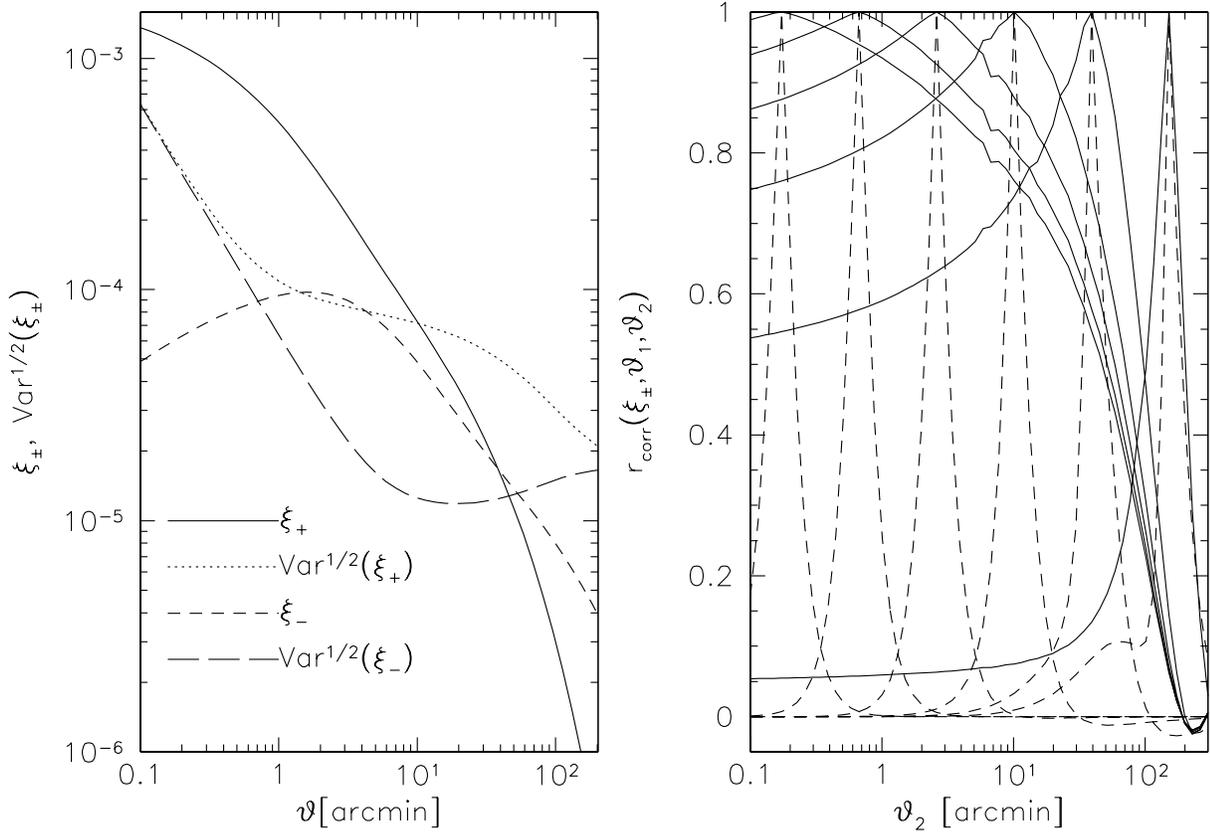}
      \caption{{\it Left panel:} The square root of the variances
   $\sqrt{{\rm Var}(\hat\xi_+;\vt)}$ and $\sqrt{{\rm Var}(\hat\xi_-;\vt)}$
   shown as dotted and long-dashed curves, together with the
   correlation functions $\xi_+(\vt)$ and $\xi_-(\vt)$ as solid and
   short-dashed curves, respectively. The model parameters are as
   described in the text; in particular, a fiducial value of the
   survey area of $1\,{\rm deg}^2$ has been taken.  For the diagonal part
   of the covariance matrix, we have assumed a relative bin size of
   $\Delta\vt/\vt=0.1$. For small $\vt$, the variance behaves as
   $\vt^{-1}$, as it is dominated by the noise from the intrinsic
   ellipticity of the source galaxies, i.e. the term $D$
   (\ref{eq:Ddef}), whereas for larger values of $\vt$, the main
   contribution comes from cosmic variance. {\it Right panel:} The
   correlation coefficient $r_{\rm corr}$, as defined in
   (\ref{eq:rcorrxi}), as a function of $\vt_2$, for various values of
   $\vt_1$. Solid curves show $r_{\rm corr}(\hat\xi_+)$, dashed
   curves show $r_{\rm corr}(\hat\xi_-)$. The value of $\vt_1$
   corresponding to each curve can be read off from the point where a
   curve attains the value $r_{\rm corr}=1$. 
              }
         \label{fig1}
   \end{figure}

Figure\ts\ref{fig1} displays in the left panel the square root of the
(auto)variance of $\hat\xi_+(\vt)$ and $\hat\xi_-(\vt)$, with ${\rm
Var}(\hat\xi_\pm;\vt)={\rm Cov}(\hat\xi_\pm,\vt;\hat\xi_\pm,\vt)$. To
calculate the value of $D$ (\ref{eq:Ddef}) which enters the diagonal
part of the covariance matrix, we have assumed a bin width of
$\Delta\vt=0.1 \vt$. The square
root of the variance -- or noise per bin -- for $\xi_+$ is smaller
than $\xi_+$ for $\vt\lesssim 10'$, and larger for larger angles (for
the assumed value of $A=1\,{\rm deg}^2$), whereas the noise for
$\xi_-$ is smaller than $\xi_-$ in an interval of $1'\lesssim
\vt\lesssim 30'$. The determination of $\xi_-$ on small angular scales
is much more difficult than for $\xi_+$, owing to the smallness of
$\xi_-$ for small $\vt$. Note that the noise scales like $A^{-1/2}$,
so that from a survey of $16\,{\rm sq.deg.}$, like the DESCART survey
(see van Waerbeke et al.\ 2001) one should be able to obtain reliable
measurements of $\xi_+$ for $\vt\lesssim 1^\circ$, and of $\xi_-$ for
$10''\lesssim \vt\lesssim 2^\circ$ in bins of relative width of
0.1. Of course, the covariance of the shear will not only depend on
the survey area, but also on its geometry; one might therefore design
survey geometries which yield the desired noise behaviour as a function
of angular scale (see, e.g., Kaiser 1998).

In order to show how strongly the correlation estimators at two
angular scales are correlated, we define the correlation coefficient
\be
r_{\rm corr}(\hat\xi_\pm;\vt_1,\vt_2)
:={{\rm Cov'}(\hat\xi_\pm,\vt_1;\hat\xi_\pm,\vt_2)\over
\sqrt{{\rm Cov'}(\hat\xi_\pm,\vt_1;\hat\xi_\pm,\vt_1)\;{\rm
Cov'}(\hat\xi_\pm,\vt_2;\hat\xi_\pm,\vt_2)}} \;,
\elabel{rcorrxi}
\ee
which is unity for $\vt_1=\vt_2$. Here, the prime indicates that the
first term in the covariances [the one proportional to
$\bar\delta(\vt_1-\vt_2)$] has been subtracted off, in order to show
the correlation induced by cosmic shear. The correlation coefficient
is plotted in the right panel of Fig.\ts\ref{fig1}. The $\xi_-$
correlation function decorrelates quickly: once the ratio between the
angular scales is larger than $\sim 2$, the correlation coefficient
(\ref{eq:rcorrxi}) has decreased to less than 10\%. In contrast to
this, the correlation function $\xi_+$ is correlated over much larger
angular scales. This was expected, given that the filter function
which relates the correlation function to the power spectrum is much
broader for $\xi_+$ than for $\xi_-$.

\def\M{\mathcal M}
\section{Covariance of the aperture mass estimator}
Due to the particular interest in the aperture mass dispersion, we
shall consider here the covariance matrix of the estimator
(\ref{eq:N23}),
\be
{\rm Cov}(\M;\theta_1,\theta_2)=\ave{\M(\theta_1)\,\M(\theta_2)}-
\ave{M_{\rm ap}^2(\theta_1)}\ave{M_{\rm ap}^2(\theta_2)}\;.
\ee
Inserting (\ref{eq:N23}) into this expression, the covariance matrix
of $\M$ becomes
\bea
{\rm Cov}(\M;\theta_1,\theta_2)\!\!&=&\!\!
{(\Delta\vt)^2\over \theta_1^2\theta_2^2}
\sum_{i=1}^{2m_1} \sum_{j=1}^{2m_2}\vt_i \vt_j
\Biggl\{K_+^2 T_+\!\rund{\vt_i\over \theta_1}T_+\!\rund{\vt_j\over \theta_2}
C_{++}(\vt_i,\vt_j)+(1-K_+)^2 T_-\!\rund{\vt_i\over \theta_1}
T_-\!\rund{\vt_j\over \theta_2} C_{--}(\vt_i,\vt_j)
\nonumber \\
&+&\!\!K_+(1-K_+) 
\eck{T_+\!\rund{\vt_i\over \theta_1}T_-\!\rund{\vt_j\over \theta_2}
C_{+-}(\vt_i,\vt_j)+ T_-\!\rund{\vt_i\over \theta_1}
T_+\!\rund{\vt_j\over \theta_2} C_{+-}(\vt_j,\vt_i)}\Biggr\} \;,
\elabel{M-1}
\eea
where we used the abbreviated notation $C_{++}(\vt_1,\vt_2)\equiv {\rm
Cov}(\hat\xi_+,\vt_1;\hat\xi_+,\vt_2)$ etc.. The values of the
aperture radius are taken to be $\theta_k=m_k\,\Delta\vt$ for $k=1,2$,
and $\vt_i=(i-1/2)\Delta\vt$, as before.  Hence, the covariance of the
aperture mass dispersion can be obtained directly in terms of the
covariances of the shear correlation functions. We can also obtain the
ensemble-averaged covariance matrix; for this purpose, we separate the
`delta-function' term in $C_{++}$ and $C_{--}$ from the rest and thus
write $C_{++}(\vt_1,\vt_2)
=D\,\bar\delta(\vt_1-\vt_2)+C'_{++}(\vt_1,\vt_2)$, similarily for
$C_{--}$.  Then the ensemble-average of (\ref{eq:M-1}) becomes
\bea
{\rm E}\rund{{\rm Cov}(\M;\theta_1,\theta_2)}\!\!&=&\!\!
{\sigma_\eps^4\over 2\pi A n^2}
\int_0^{2{\rm min}(\theta_1,\theta_2)}
{\d\vt\,\vt\over \theta_1^2\theta_2^2}
\eck{K_+^2 T_+\!\rund{\vt\over\theta_1}T_+\!\rund{\vt\over\theta_2}
+(1-K_+)^2 T_-\!\rund{\vt\over\theta_1}T_-\!\rund{\vt\over\theta_2}}
\nonumber \\
+\int_0^{2\theta_1}\!{\d\vt_1\,\vt_1\over\theta_1^2}
\!\!\!\!&&\!\!\!\!\!
\int_0^{2\theta_2}\!{\d\vt_2\,\vt_2\over\theta_2^2}
\Biggl\{K_+^2 T_+\!\rund{\vt_1\over \theta_1}T_+\!\rund{\vt_2\over \theta_2}
C'_{++}(\vt_1,\vt_2)+(1-K_+)^2 T_-\!\rund{\vt_1\over \theta_1}
T_-\!\rund{\vt_2\over \theta_2} C'_{--}(\vt_1,\vt_2)
\nonumber\\
&+&\!\!K_+(1-K_+) 
\eck{T_+\!\rund{\vt_1\over \theta_1}T_-\!\rund{\vt_2\over \theta_2}
C_{+-}(\vt_1,\vt_2)+ T_-\!\rund{\vt_1\over \theta_1}
T_+\!\rund{\vt_2\over \theta_2} C_{+-}(\vt_2,\vt_1)}\Biggr\} \;.
\elabel{M-2}
\eea
As expected, the covariance matrix of $\M$ does not depend on the
binning of the correlation function. For the calculation of $\M$ for
observational data, it is therefore best to choose very small bin
widths, in order to minimize discretization errors.

The first term in (\ref{eq:M-2}) yields the covariance in the absence
of cosmic shear correlations, i.e. the covariance of the estimator
$\M$ due to the intrinsic ellipticity of the source galaxies. This
term can be written as 
\[
{\sigma_\eps^4\over 2\pi A n^2 \theta_2^2}\,f(\theta_1/\theta_2)
=3.939\times 10^{-10}\rund{\sigma_\eps\over 0.3}^4\rund{A\over 1\,{\rm
deg}^2}^{-1} \rund{n\over 30\,{\rm
arcmin}^{-2}}^{-2}\rund{\theta_2\over 1\,{\rm
arcmin}}^{-2}\;f(\theta_1/\theta_2) \;,
\]
where
\bea
f(R)&=&(1-2K_++2K_+^2)\int_0^2\d x\,x\, T_+(x)T_+(R x)
=(1-2K_++2K_+^2)\int_0^2\d x\,x\,T_-(x)T_-(Rx)\nonumber \\
&=&(1-2K_++2K_+^2)(576)^2\int_0^\infty{\d t\over R^4t^7}
{\rm J}_4^2(Rt)\,{\rm J}_4^2(t)
\eea
where the explicit relation has been obtained by making use of the
original definitions of $T_\pm$. The function
$f(R)$ is maximized at $R=1$, where its value is $f(1)\approx 0.29$,
and it decreases quickly for appreciable ratios of the aperture
radii. The dependence on $K_+$ is simple, and obviously this noise
term is minimized for $K_+=1/2$, i.e. when both correlation functions
enter the estimate $\M$ with equal weight. The
variance of the estimator $\M$ in the absence of shear correlations
can be compared with the corresponding expression obtained in SvWJK,
obtained from their equations (5.12) and (5.16), yielding
\[
{G^2\,\sigma_\eps^4\over 2\pi^2 n^2\theta^4 N_{\rm f}}
={2.88\over \pi^2}{\sigma_\eps^4\over A n^2 \theta^2}\;,
\]
where $G=1.2$ is a numerical coefficient (calculated from the filter
function in the definition of $M_{\rm ap}$), and $N_{\rm f}$ is the
number of independent apertures that can be placed on the data field,
taken to be $A/(4\theta^2)$. Together with $f(1)\approx0.29$, one
clearly sees that the variance of the estimator $\M$ is substantially
smaller than that used in SvWJK which was based on placing
non-overlapping apertures
on the data field. The main reason for this difference is that in the
latter method, only one component of the source ellipticities is used;
furthermore, apertures separated by less than their diameter are
statistically nearly independent. Hence, the determination of
$\ave{M_{\rm ap}^2}$ through the shear correlation function is not
only more practical in the presence of gaps in the data field, but
also far more efficient than the alternative method.

   \begin{figure}
   \centering
   \includegraphics[width=17cm]{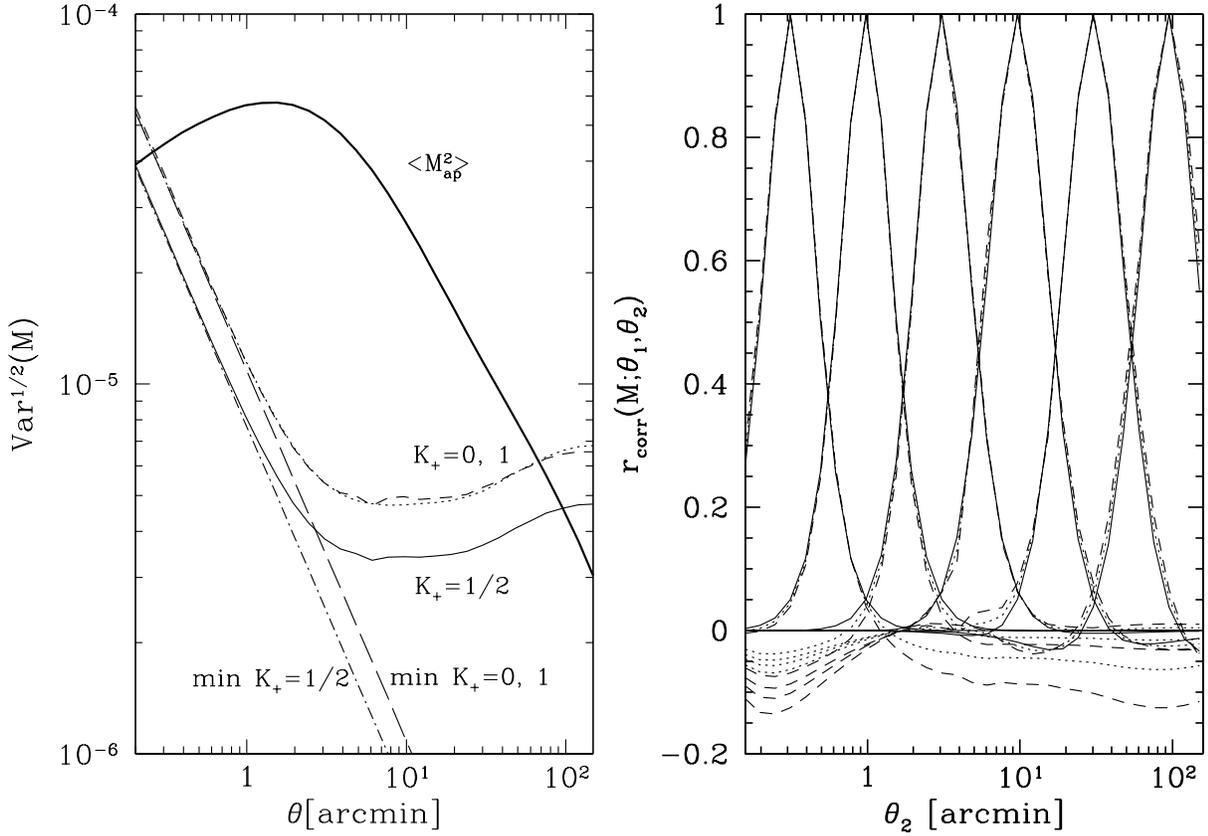}
      \caption{
   {\it Left panel:} The square-root of the autovariance of
   $\M$ as a function of angular scale. The long-dashed and
   dash-dotted curves show the minimum variance, i.e. in the absence
   of a shear correlation; this variance is due solely to the
   intrinsic ellipticity of source galaxies. Curves are shown for
   $K_+=0, 1/2, 1$, where the minimum variance is the same for $K_+=0$
   and $1$. The solid, dashed and dotted curves show the variance
   in the presence of a cosmic shear; also here, the cases $K_+=0$ and
   $K_-=1$ are nearly the same, and the variance is smallest for
   $K_+=1/2$. For comparison, $\ave{M_{\rm ap}^2}(\theta)$ is plotted
   as thick solid curve. As for the other figures shown before, our
   standard set of parameters $A=1\,{\rm deg}^2$, $n=30\,{\rm
   arcmin}^{-2}$ and $\sigma_\eps=0.3$ has been used; the variance
   scales as $A^{-1}$. {\it Right panel:} The correlation coefficient
   $r_{\rm corr}(\M;\theta_1,\theta_2)$ of the covariance of the
   estimator $\M$ is plotted as a function of $\theta_2$, for various
   values of $\theta_1$; the values of $\theta_1$ can be localized as
   those points where the correlation function attains the value
   unity. The solid curves are for $K_+=0$, i.e. when only the
   correlation function $\xi_-$ is used in the estimate of $\M$, the
   dotted curves are for $K_+=1/2$, and the dashed curves for
   $K_+=1$. The width of all three families of curves is very similar
   and (in logarithmic terms) basically independent of $\theta_1$. The
   $K_+=0$ curves do not develop a tail of anticorrelation, as is the
   case for $K_+=1$ (and therefore also for $K_+=1/2$). Hence, whereas
   $K_+=1/2$ yields the smallest variance of the estimator $\M$, it
   leads to a small but long-range correlation between different
   angular scales
              }
         \label{fig8}
   \end{figure}

We have plotted in Fig.\ts\ref{fig8} the square root of the variance
${\rm Var}(\M,\theta)\equiv {\rm Cov}(\M;\theta,\theta)$ of $\M$, both
for the case of no correlations, and for our standard model for the
cosmic shear. For small angular scales, the variance is completely
dominated by the intrinsic source ellipticities, whereas the cosmic
variance is the important noise error for larger angular scales. For
the parameters used here, the transition between these two regimes
occurs at a few arcminutes. It must be noted that the shape of the
variance curves are independent of the survey area $A$. The results
shown in Fig.\ts\ref{fig8} can be seen as an a posteriori
justification of using the Gaussian assumption for the calculation of
the shear four-point function in Sect.\ts 4. In Fig.\ts 4 of van
Waerbeke et al.\ts (2002), the influence of the non-linear density
evolution on the kurtosis of $M_{\rm ap}$ was studied, using
ray-tracing simulations through numerically generated cosmological
matter distributions. The non-Gaussian effects start to become
non-negligible for angular scales below $\sim 5'$. As can be seen in
Fig.\ts\ref{fig8}, this is about the scale where the transition occurs
between the variance being dominated by the intrinsic ellipticity
distribution and the cosmic variance. Hence, we can expect that a more
advanced treatment of the shear four-point function would yield a
slightly larger variance in this transition region around $\sim 5'$:
for significantly smaller scales, it is dominated by the intrinsic
ellipticity noise, and for larger scales, the shear four-point
function is basically Gaussian.

The variances of $\M$ for the cases $K_+=0$ and $K_+=1$ are basically
identical, and larger by a factor $\sim {2}$ than the variance of the
estimator for $K_+=1/2$. Hence, to minimize the variance of the
estimator $\M$, $K_+=1/2$ should be chosen. With this choice, the
results are unchanged even in the presence of a B-mode contribution
(see SvWM).  As was already mentioned by C02, using cosmic shear
estimators which use $\xi_+$ and $\xi_-$ with equal weight reduces the
resulting noise by a factor $2^{-1/2}$. One notes that the variance
for large $\theta$ rises, but very slowly. We can compare the
behavior of the variance of $\M$ with that derived in SvWJK for a
more direct estimator for the aperture mass dispersion; using eqs.\ts
(5.12) and (5.16) of that paper, one finds in the limit of large
angles (and, to make the estimate comparable to the one obtained here,
zero kurtosis) that $\sqrt{{\rm Var}(\M';\theta)}\approx
{2\theta\ave{M_{\rm ap}^2}(\theta)/ A}$, where $\M'$ is the estimator
used in SvWJK. The functional behavior with $\theta$ is similar to
that seen in Fig.\ts\ref{fig8}, but the amplitude is lower by a factor
of about 2 for $K_+=0, 1$, and about 3 for $K_+=1/2$; this again shows
the superiority of the estimator considered here in comparison to
laying down independent apertures on the data field.

To investigate the correlation of the estimator $\M$ between different
angular scales, we define the correlation coefficient
\be
r_{\rm corr}(\M;\theta_1,\theta_2):={{\rm Cov}(\M;\theta_1,\theta_2)\over
\sqrt{{\rm Cov}(\M;\theta_1,\theta_1)\;{\rm
Cov}(\M;\theta_2,\theta_2)}} \;,
\elabel{rcorrM}
\ee
which has the property that $r_{\rm corr}(\M;\theta,\theta)=1$. The
dependence of this correlation on the ratio of the angular scales then
provides information on the correlated error of the determination of
the aperture mass dispersion on different angular scales. In the right
panel of Fig.\ts\ref{fig8} we have plotted this correlation
coefficient as a function of $\theta_2$, for various values of
$\theta_1$. The logarithmic representation clearly shows that this
correlation coefficient depends mainly on the ratio
$\theta_1/\theta_2$. The correlation drops off quickly, so that
estimates of $\ave{M_{\rm ap}^2}(\theta)$ for two angles with ratio
$\theta_1/\theta_2\lesssim 1/3$ or $\theta_1/\theta_2 \gtrsim 3$ are
practically uncorrelated. This was to be expected given that
$\ave{M_{\rm ap}^2}(\theta)$ is obtained from the power spectrum
$P_\kappa(\ell)$ through a very well localized filter function. Hence,
the estimator $\M$ decorrelates much faster than those of the shear
correlation functions. Also seen in Fig.\ts\ref{fig8} is the fact that
in the case of $K_+=1/2$ and $K_+=1$, the correlation coefficient
attains long negative, but low-amplitude tails, whereas they are
basically absent if $K_+=0$. This is due to the much faster
decorrelation of $\xi_-$ with scale ratio than that of $\xi_+$.

\section{A simple estimator for the power spectrum, and its
covariance}
The relations (\ref{eq:7}) may be used to define an estimator for the
power spectrum $P_\kappa(\ell)$ as
\be
\hat P(\ell)=2\pi\int_{\theta_{\rm min}}^{\theta_{\rm max}}
\d\theta\;\theta\,\eck{K_+\xi_+(\theta){\rm J}_0(\ell\theta)
+(1-K_+)\xi_-(\theta){\rm J}_4(\ell\theta)}\;,
\elabel{powest}
\ee
where $K_+\in[0,1]$ again describes the relative contribution from the
$\xi_+$ correlation. Here, $\theta_{\rm min}$ and $\theta_{\rm max}$
describe the range over which the correlation function has been
measured. If this range extends from zero to infinity, the estimator
(\ref{eq:powest}) would be unbiased (and would yield the E-mode power
spectrum for $K_+=1/2$ even in the presence of B-modes); for real
datasets, where this range is finite, (\ref{eq:powest}) is a biased
estimator. Note that in the absence of B-modes, eq.\ts
(\ref{eq:powest}) remains valid even if the factor $K_+$ is chosen to
be a function of $\ell$. The expectation value can be obtained by
inserting the relation (\ref{eq:2}) between the correlation functions
and the true power spectrum into (\ref{eq:powest}) to yield
\be
\ave{\hat P(\ell)}\equiv P_{\rm obs}(\ell)
=\int_0^\infty\d\ell'\;\ell'\,\eck{K_+
G_0(\ell,\ell')+(1-K_+)G_4(\ell,\ell')}\;P_\kappa(\ell') \;,
\elabel{Pobs}
\ee 
with
\be
G_n(\ell,\ell')=\int_{\theta_{\rm min}}^{\theta_{\rm max}}
\d\theta\;\theta\,{\rm J}_n(\ell\theta)\,{\rm J}_n(\ell'\theta)
=\eck{{\theta\over \ell'^2-\ell^2}
\wave{\ell' {\rm J}_{n+1}(\ell'\theta){\rm J}_n(\ell\theta)
-\ell {\rm J}_n(\ell'\theta){\rm
 J}_{n+1}(\ell\theta)}}^{\theta=\theta_{\rm max}}_{\theta=\theta_{\rm
 min}}
\;. \nonumber
\ee

   \begin{figure}
   \centering
   \includegraphics[width=13cm]{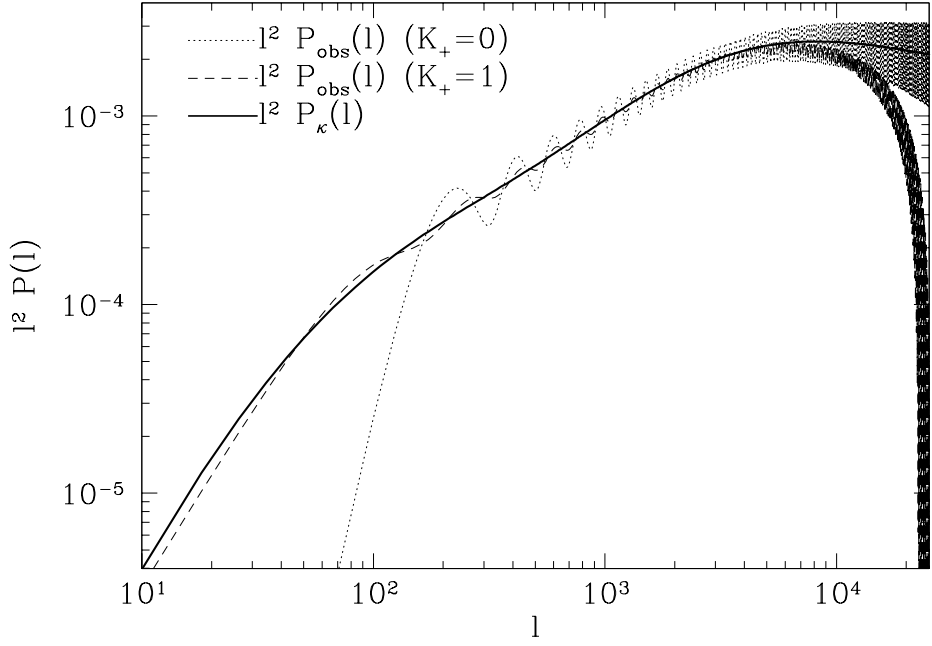}
      \caption{The thick solid line displays the dimensionless projected
   power spectrum $\ell^2 \P_\kappa(\ell)$, whereas the other two
   curves show the `observed' power spectrum, as defined in
   (\ref{eq:Pobs}). The dotted curve is for $K_+=0$, i.e. only $\xi_-$
   enters the determination of the observed power spectrum in this
   case; the dashed curve is for $K_+=1$. In this plot is was assumed
   that the correlation functions are known between $\theta_{\rm
   min}=6''$ and $\theta_{\rm min}=2^\circ$   
              }
         \label{fig5}
   \end{figure}

We have plotted the `observed' power spectrum as $\ell^2 P_{\rm
obs}(\ell)$ in Fig.\ts\ref{fig5}, for $K_+=0$ and $K_+=1$, assuming
that $\theta_{\rm min}=6''$ and $\theta_{\rm max}=2^\circ$. A
comparison with the underlying power spectrum (shown as heavy solid
curve) shows that $P_{\rm obs}$ traces the true power spectrum over a
wide range of $\ell$-values, though in an oscillatory way. If $P_{\rm
obs}$ is determined solely from $\xi_-$ (i.e. $K_+=0$), it
substantially underestimates the power for $\ell \lesssim 10^2$ (that
is, approximately for $\ell\lesssim 2\pi/\theta_{\rm max}$), but
traces the true power spectrum out to the largest values of $\ell$
plotted. Conversely, the observed power determined from $\xi_+$ yields
good estimates of the true power for small values of $\ell$, but
deviates from it strongly for $\ell\gtrsim 8\times 10^3$, that is for
values of $\ell$ much less than $2\pi/\theta_{\rm min}\sim 2\times
10^5$. The different behavior of the two estimates again is due to the
different filter function through which correlation function and power
spectrum are related. Fig.\ts\ref{fig5} suggests that the best
estimate for the power spectrum is obtained by choosing $K_+=1$ for
small values of $\ell$, and $K_+=0$ for large $\ell$. 

The covariance matrix of $\hat P$ reads, for $K_+=1$,
\be
{\rm Cov}(\hat P;\ell,\ell')
=(2\pi)^2\int_{\theta_{\rm min}}^{\theta_{\rm max}}
\d\theta\;\theta\,{\rm J}_0(\ell\theta)
\int_{\theta_{\rm min}}^{\theta_{\rm max}}
\d\theta'\;\theta'\,{\rm J}_0(\ell'\theta')\,C_{++}(\theta,\theta')\;;
\nonumber
\ee
the generalization for other values of $K_+$ is obvious and shall not
be reproduced here.

To estimate the power spectrum from cosmic shear data, it is useful to
define the power in a band with upper and lower $\ell$-values
$\ell_{i{\rm u}}$ and $\ell_{i{\rm l}}$ as
\be
\P_i:={1\over \Delta_i}
\int_{\ell_{i{\rm l}}}^{\ell_{i{\rm u}}}\d\ell\;\ell\,\hat
P(\ell)
={2\pi\over \Delta_i}\int_{\theta_{\rm min}}^{\theta_{\rm max}}
{\d\theta\over\theta}
\wave{K_+ \xi_+(\theta) \Bigl[g_+(\ell_{i{\rm u}}\theta)-g_+(\ell_{i{\rm
l}}\theta)\Bigr]
+(1-K_+) \xi_-(\theta) \Bigl[g_-(\ell_{i{\rm u}}\theta)-g_-(\ell_{i{\rm
l}}\theta)\Bigr]}\;,
\ee
where $\Delta_i=\ln(\ell_{i{\rm u}}/\ell_{i{\rm l}})$ is the
logarithmic width of the band, and
\be
g_+(x)=x{\rm J}_1(x)\quad ;\quad
g_-(x)=\rund{x-{8\over x}}{\rm J}_1(x)-8{\rm J}_2(x)\;.
\ee
One expects that the band power traces $\bar\ell_i^2
\,P_\kappa(\bar\ell_i)$, where $\bar\ell_i$ is the geometric mean of
$\ell_{i{\rm u}}$ and $\ell_{i{\rm l}}$, i.e. the center of the bin.
The covariance of the band power of two bins $i$ and $j$ is
\bea
{\rm Cov}(\P_i\P_j)&=&{2\pi\sigma_\eps^4\over \Delta_i\,\Delta_j\,A\,n^2}
\int_{\theta_{\rm min}}^{\theta_{\rm max}}{\d\theta\over\theta^3}
\Bigl[ K_+^2\,G_{i+}(\theta)G_{j+}(\theta)
+(1-K_+)^2G_{i-}(\theta)G_{j-}(\theta)\Bigr] 
\nonumber \\
&+& {(2\pi)^2\over \Delta_i\,\Delta_j}
\int_{\theta_{\rm min}}^{\theta_{\rm max}}{\d\theta\over\theta}
\int_{\theta_{\rm min}}^{\theta_{\rm max}}{\d\theta'\over\theta'}
\Bigl\{K_+^2\,C'_{++}(\theta,\theta')\,G_{i+}(\theta)G_{j+}(\theta')
+(1-K_+)^2 C'_{--}(\theta,\theta')\,G_{i-}(\theta)G_{j-}(\theta')
\nonumber \\
&+&K_+(1-K_+)C_{+-}(\theta,\theta')\eck{G_{i+}(\theta)G_{j-}(\theta')
+G_{i-}(\theta')G_{j+}(\theta)}
\Bigr\} 
\elabel{CovPP}
\eea
where $G_{i\pm}(\theta)=g_\pm(\ell_{i{\rm u}}\theta)-g_\pm(\ell_{i{\rm
l}}\theta)$. In (\ref{eq:CovPP}), we have already split off the
`delta-function' part of the correlation covariance matrices, which
yields the first term. It should be noted that the foregoing
expressions remain valid if $K_+$ is varied as a function of
$\ell$. From Fig.\ts\ref{fig8} it is clear that in order to get the
least bias, one wants to choose $K_+\sim 1$ for small $\ell$, and
$K_+\sim 0$ for large $\ell$; for the intermediate region, setting
$K_+=1/2$ should yield the smallest error on the power spectrum. We
have therefore constructed a function $K_+(\ell)$ which has these 
desired properties.

   \begin{figure}
   \centering
   \includegraphics[width=14cm]{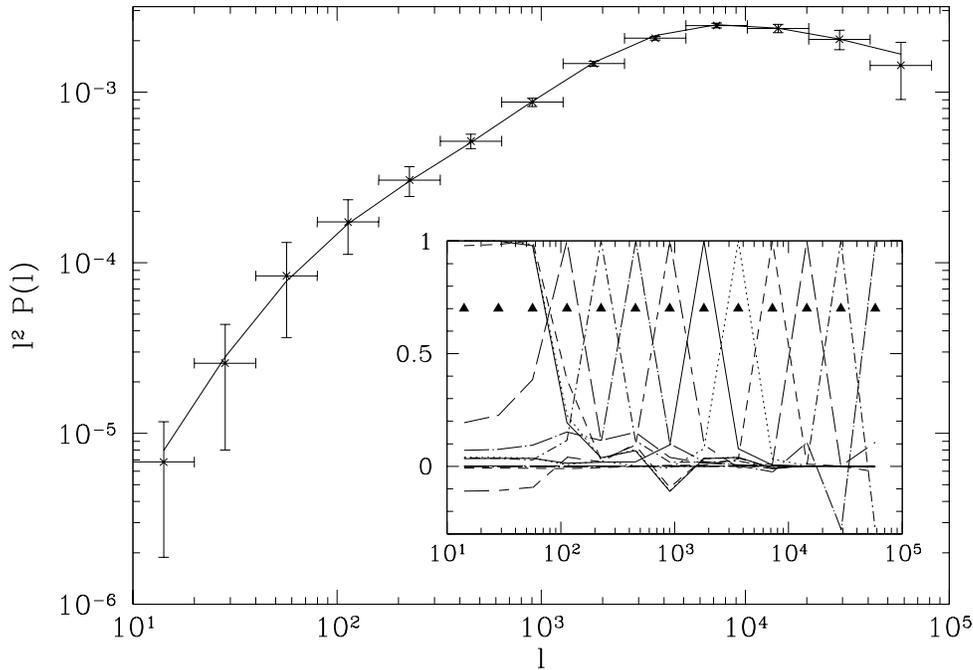}
      \caption{The large panel shows the estimates of the band power,
   shown as horizontal bars whose length indicates the bins used. The
   error bar on each bin shows the square root of the autovariance of
   the band power, and the solid curve is the underlying power
   spectrum, $\ell^2 P_\kappa(\ell)$. For this figure, we have assumed
   that the correlation functions are measured for $6''\le\vt\le
   2^\circ$, from a survey of $A=25\,{\deg}^2$. The inset figure shows
   the correlation coefficient between the 13 different bins, where
   the triangles indicate the center $\bar\ell$ of each bin. One sees
   that the bands are very little correlated, except for the three
   bins with smallest $\ell$; in fact, the first three band power
   estimates are fully correlated. This explains why the band-power
   estimator yields reasonable results even for $\ell <
   2\pi/\theta_{\rm max}\sim 180$ -- this is just a coincidence.
              }
         \label{fig9}
   \end{figure}

In Fig.\ts\ref{fig9} we have plotted the band power for our reference
model parameters, in 13 bins of width $\ell_{i{\rm u}}/\ell_{i{\rm
l}}=2$, between $\ell=10$ and $\ell\approx 8\times 10^4$. The band
power is shown as crosses, and vertical error bars show the range of
the bins. For comparison, the solid curve shows
$\ell^2\,P_\kappa(\ell)$; as expected, with this new choice of
$K_+(\ell)$, the band power traces the underlying power spectrum over
a very wide range of wavenumbers. Only in the bins with the smallest
and largest value of $\ell$ is there a significant bias; over the
range $2\pi/\theta_{\rm max}\approx 180\lesssim\ell\lesssim
2\pi/\theta_{\rm min}\approx 2\times 10^5$, the band power estimator
is practically unbiased.  Next we calculated the error bars on the
band power, by taking the square root of the diagonal part of
(\ref{eq:CovPP}). For this calculation, we have assumed to have a
total area of $A=25\,{\rm deg}^2$, for which the condition $\theta_{\rm
max}^2\ll A$ for the validity of the treatment of the ensemble average
in Sect.\ts 5 is approximately satisfied. The square root of this
autovariance is plotted as errorbars on the band power in
Fig.\ts\ref{fig9}; as can be seen from this figure, the
signal-to-noise ratio is larger than unity in all bins shown, and in
fact very large for intermediate values of $\ell$. Hence, the power
spectrum $P_\kappa(\ell)$ can be measured over a broad range of $\ell$
for the parameters chosen here.

Of course, in order to interpret the error bars correctly, it is
important to see the degree of correlated noise between different
bands. The correlation matrix for the bins [defined in full analogy to
(\ref{eq:rcorrM})] was calculated and its values are plotted in the
inset of Fig.\ts\ref{fig9}. One sees that errors of the bins for
intermediate and high values of $\ell$ are essentially uncorrelated
(the correlation coefficient for neighboring bins is $\lesssim 10\%$
for $\ell\gtrsim 200$); however, for $\ell\lesssim 100$ the bins
become strongly correlated. In fact, the agreement of the band powers
with the underlying power spectrum is forticious for $\ell \lesssim
100$: the three band powers at lowest $\ell$ are nearly fully
correlated, so that these three points contain practically the same
information of the underlying power spectrum.

The method presented here for the determination of the power spectrum
has the virtue of its simplicity. Other methods for determining the
power spectrum from shear data have been investigated, e.g. by Kaiser
(1998), Seljak (1998) and Hu \& White (2001). Our approach has the
property that it makes use only of the shear correlation functions,
not on the spatial distribution of the shear. Since the shear
correlation function contains all two-point statistical information of
the shear field, no information loss occurs. Comparing the results of
Fig.\ts\ref{fig9} with those of Hu \& White (2001) it seems that both
methods yield very similar error bars of the power spectrum, and that
in the respective $\ell$-range of applicability, the decorrelation
between neighboring bins is equally quick. Since our method does not
require the `pixelization' of shear data, it can estimate the power
spectrum to larger values of $\ell$.

\section{Constraints on cosmological parameters}
One of the central goals of cosmic shear research is the determination
of cosmological parameters. Since the power spectrum $P_\kappa$, and
thus the shear correlation functions, depend on the cosmological
model, precise measurements of cosmic shear can be used to tie down
the range of allowable model parameters (e.g., Jain \& Seljak 1997;
Bernardeau et al.\ 1997). The largest cosmic shear survey today
already yielded significant model constraints (van Waerbeke et al.\
2001, 2002; Hoekstra et al.\ 2002). We shall briefly discuss the
expected confidence regions in parameter space, using the previously
calculated covariance matrix of the shear correlation functions.
For that, we consider a figure-of-merit function
\be
\chi^2(p):=\sum_{ij}\rund{\xi_i(p)-\xi_i^{\rm t}}{\rm Cov}_{ij}^{-1}
\rund{\xi_j(p)-\xi_j^{\rm t}}\;,
\elabel{chisq}
\ee
where the index $^{\rm t}$ indicates the correlation function of the
input model, $p$ is a set of model parameters, and the summation indices
label the angular bins of the correlation function. We shall consider
three different kinds of the function (\ref{eq:chisq}): in the first,
named $\chi^2_+$ hereafter, the correlation functions in
(\ref{eq:chisq}) are the $\xi_+$ correlations, and the covariance
matrix corresponds to (\ref{eq:Cpp}). The second kind is denoted by
$\chi^2_-$ and uses the $\xi_-$ correlation function. The third kind,
$\chi^2_{\rm tot}$, is obtained by constructing a vector
$\xi_i=(\xi_{+1},\xi_{+2},\dots,
\xi_{+N},\xi_{-1},\xi_{-2},\dots,\xi_{-N})$, when there are $N$ angular
bins in which $\xi_\pm$ has been measured. Correspondingly, the
covariance matrix in (\ref{eq:chisq}) is the 
$2\times2$-block matrix with $C_{++}$ and $C_{--}$ in the upper left
and lower right quadrant, respectively, and $C_{+-}$ in the other two.

   \begin{figure}
   \centering
   \includegraphics[width=14cm]{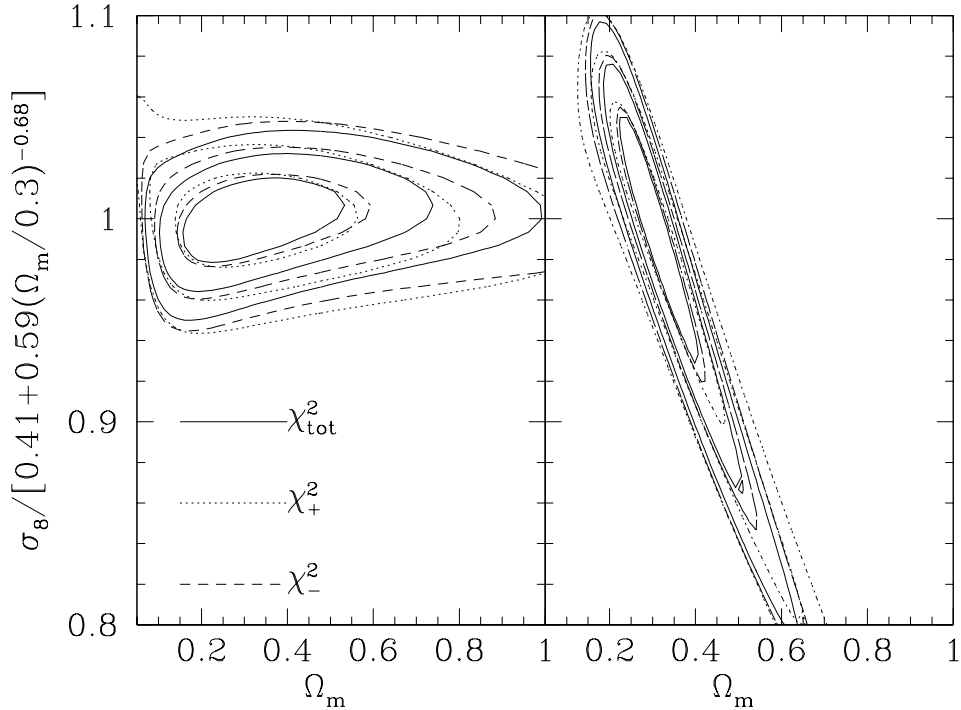}
      \caption{1-$\sigma$, 2-$\sigma$ and 3-$\sigma$ confidence
   contours in the $\Omega_{\rm m}$-$\Sigma_8$-plane, where $\Sigma_8$
   is a scaled version of the power spectrum normalization parameter
   $\sigma_8$, as indicated. Dotted, dashed and solid contours
   correspond to $\chi^2_+$, $\chi^2_-$ and $\chi^2_{\rm tot}$. In the
   left panel, the shape parameter of the power spectrum
   $\Gamma=0.21$, whereas in the right panel, $\Gamma=0.7\Omega_{\rm
   m}$. The reference model is the one used before, i.e. $\Omega_{\rm
   m}=0.3$, $\Omega_\Lambda=1-\Omega_{\rm m}$, $\sigma_8=1$. The
   contours are obtained by assuming a survey area of $A=5\,{\rm
   deg}^2$, and that the correlation functions were measured in the
   range $12''\le\vt\le 30'$. 
              }
         \label{fig4}
   \end{figure}

\subsection{The $\sigma_8$--$\Omega_{\rm m}$ degeneracy}
We first have calculated $\chi^2$ for a two-parameter set of cosmological
models, by varying $\Omega_{\rm m}$ and $\sigma_8$; these two
parameters were considered before in van Waerbeke et al.\ (2001). We
kept the other cosmological parameters fixed, except that flat
Universes are considered, i.e. $\Omega_{\rm m}+\Omega_\Lambda=1$. From
the confidence contours plotted in van Waerbeke et al., it is
anticipated that the $\chi^2$-function, considered as a function of
$\Omega_{\rm m}$ and $\sigma_8$, will have a long and curved
valley close to its minimum. This was verified here. In it therefore
useful to consider the combination
$\Sigma_8:= \sigma_8\eck{0.41+0.59(\Omega_{\rm m}/0.3)^{-0.68}}^{-1}$
as parameter when plotting contours of $\chi^2$ (the numerical values
occurring in this definition have been obtained by a fit through the
valley line in the $\Omega_{\rm m}$--$\sigma_8$ plane). 

In Fig.\ts\ref{fig4} we have plotted contours of constant $\chi^2$ in
the $\Omega_{\rm m}$--$\Sigma_8$ plane, corresponding to 1-$\sigma$,
2-$\sigma$ and 3-$\sigma$ confidence regions. In the left panel, we
kept the shape parameter of the power spectrum fixed, $\Gamma=0.21$,
whereas in the right panel we used $\Gamma=0.7\Omega_{\rm m}$,
adequate for a (dimensionless) Hubble constant of $h=0.7$. Shown are
the confidence regions for all three functions $\chi^2$, as
indicated. The first point to note is that, for a given value of
$\Omega_{\rm m}$, $\sigma_8$ is very well
constrained, to within a few percent. This implies that the
normalization of the power spectrum is very well determined from
cosmic shear observations. Secondly, the 1-$\sigma$ uncertainty on
$\Omega_{\rm m}$ is about 0.1 for an assumed survey size of $A=5\,{\rm
deg}^2$; indeed, the left panel of Fig.\ts\ref{fig4} can be compared
directly with similar figures in van Waerbeke et al., and the
constraints on $\Omega_{\rm m}$ are quite similar. Third, if the shape
parameter changes with $\Omega_{\rm m}$, the confidence regions are
narrower than when setting $\Gamma=0.21$ as constant, which implies
that the shear correlation functions are sensitive measures for
$\Gamma$. 

   \begin{figure} \centering
   \includegraphics[width=20cm]{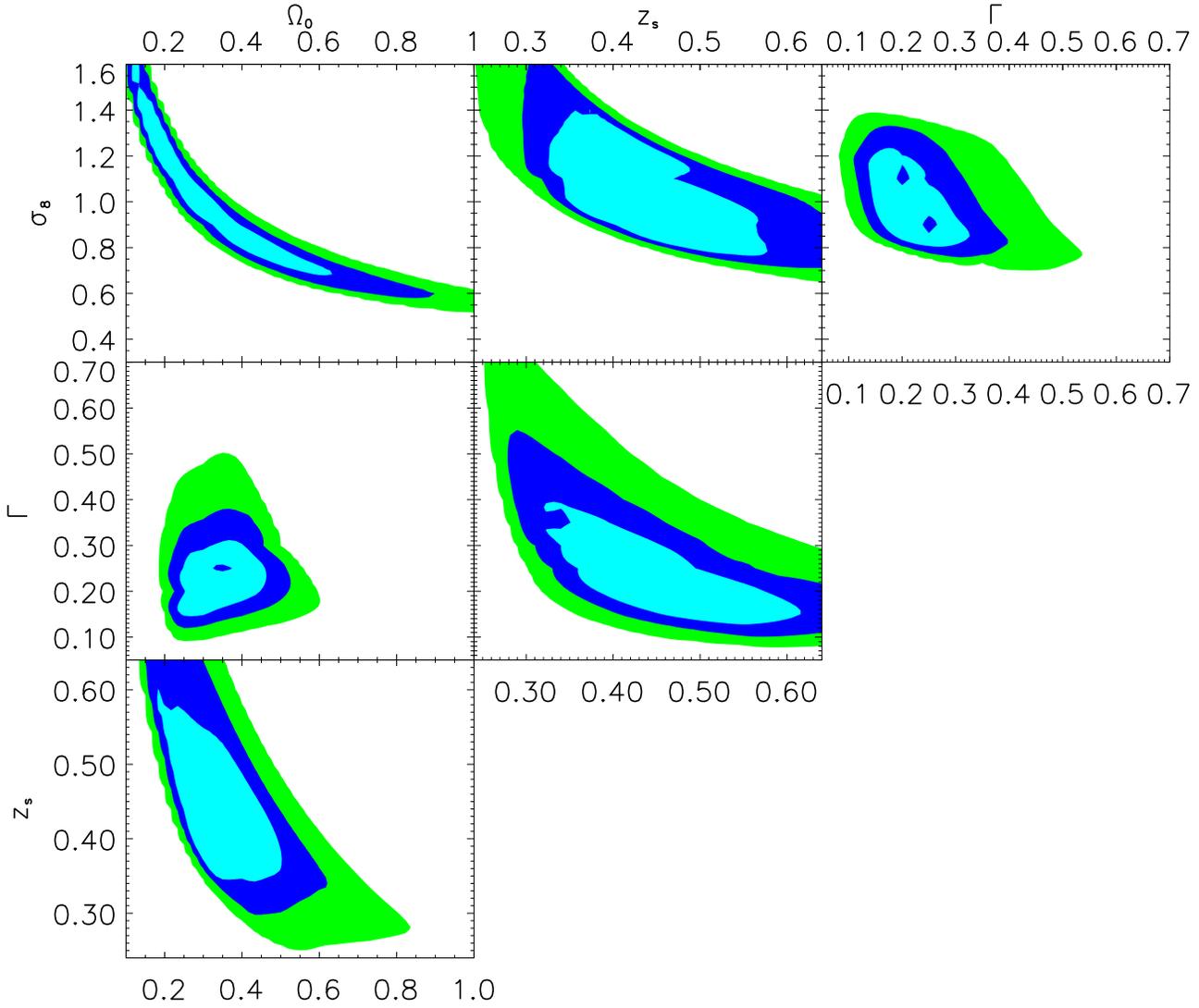} \caption{1-$\sigma$,
   2-$\sigma$ and 3-$\sigma$ confidence contours for the maximum
   likelihood analysis on the four parameters $\Omega_{\rm m}$,
   $\sigma_8$, $\Gamma$ and the source redshift parameter $z_{\rm s}$
   (see text). The six possible pairs of parameters are displayed. On
   each figure, the two hidden parameters are marginalized such that
   $\Omega_{\rm m}\in [0.2,0.4]$, $\sigma_8\in [0.8,1.1]$, $\Gamma\in
   [0.1,0.3]$ and $z_{\rm s}\in [0.4,0.5]$, and the cosmological
   constant is fixed to $\Omega_\Lambda=1-\Omega_{\rm m}$. The reference
   model is $\Omega_{\rm m}=0.3$, $\sigma_8=1$, $\Gamma=0.21$ and
   $z_{\rm s}=0.44$. The survey area is $A=16\,{\rm deg}^2$, the
   galaxy ellipticity r.m.s. is $0.3$, and the correlation functions
   are measured in the range $0\arcminf6\le\vt\le 30'$.  }
   \label{figweakprior} \end{figure}
   \begin{figure} \centering
   \includegraphics[width=20cm]{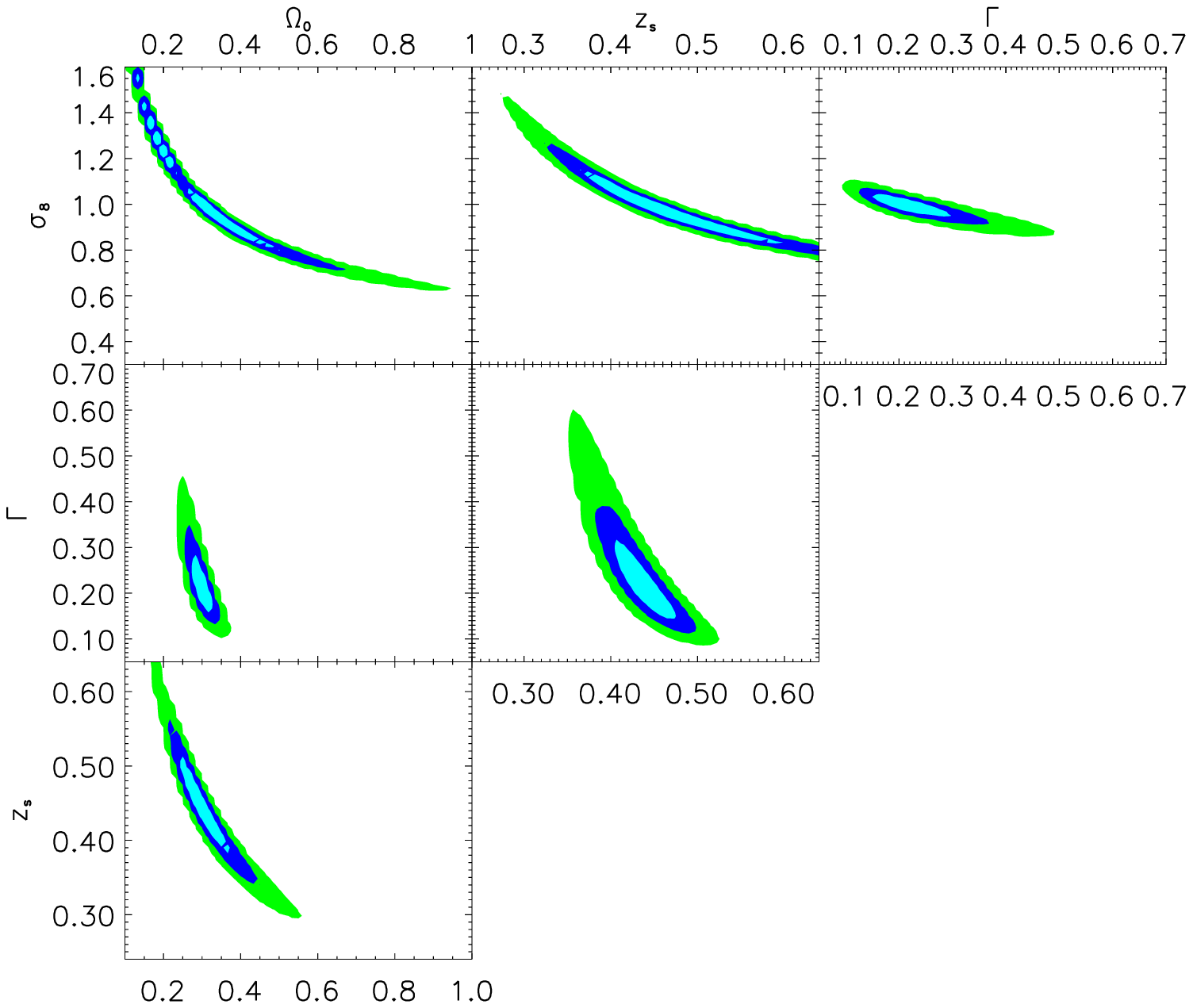} \caption{Same as
   figure \ref{figweakprior} with strong priors: in each figure, the
   two hidden parameters as assumed to be known perfectly. These plots
   show the degeneracy directions among all the possible pairs of
   parameters obtained from $\Omega_{\rm m}$, $\sigma_8$, $\Gamma$ and
   $z_{\rm s}$.  Note that the wiggles at the edge of the contours are not
   real features of the probability constraints. Their are inherent to
   the sampling limitation of the 4-dimensional cube of models given
   the memory limit of our machines.  Also note that the upper left
   panel is the analogue of Fig.\ts\ref{fig4}, but without the scaling
   employed there.
} \label{figstrongprior}
   \end{figure}

\subsection{Additional parameters}
In the prospect of measuring the cosmological parameters, we must also
allow for our lack of knowledge of some parameters which might have a
significant effect on our observable, but which are poorly
constrained. In particular, it is important to explore the possible
directions of degeneracies and to identify which combination of
parameters offers the most promising measurements. As shown in Jain \&
Seljak (1997), the shear 2-point statistics depend primarily on four
parameters: $\Omega_{\rm m}$, $\sigma_8$, $\Gamma$ and $z_{\rm s}$,
the source redshift parameter. We now model the source redshift
distribution with
\begin{equation}
p(z)={\beta\over z_{\rm s} \ \Gamma\left({1+\alpha\over \beta}\right)} 
\left({z\over
z_{\rm s}}\right)^\alpha \exp\left[-\left({z\over z_{\rm
s}}\right)^\beta\right],
\label{zsource}
\end{equation}
where we choose $\alpha=2$ and $\beta=1.2$, and take $z_{\rm s}$ as
the free parameter; note that the mean redshift of the sources is
$\ave{z}\approx 2.09 z_{\rm s}$.
This choice is motivated from observational
considerations (see van Waerbeke et al.\ 2002).  We therefore
constructed the 4-dimensional figure-of-merit function
(\ref{eq:chisq}), where now $\xi_i\equiv \xi_+(\theta_i;\Omega_{\rm
m},\sigma_8,\Gamma,z_{\rm s})$.  The fiducial model is chosen to be
$\Omega_{\rm m}=0.3$, $\sigma_8=1$, $\Gamma=0.21$ and $z_{\rm
s}=0.44$.  It is assumed that the shear correlation function is
measured in the range $0\arcminf6 \le \vt\le 30'$, comparable with the
most recent cosmic shear measurements. We also fix the survey area to
$A=16\,{\rm deg}^2$ and the galaxy intrinsic ellipticity r.m.s. to
$0.3$. Figure \ref{figweakprior} shows the confidence regions,
assuming weak priors on pairs of hidden parameters. From the four
parameters to be determined, we can construct 6 pairs of parameters,
and for each pair, the other two parameters are marginalized, in such
a way that $\Omega_{\rm m}\in [0.2,0.4]$, $\sigma_8\in [0.8,1.1]$,
$\Gamma\in [0.1,0.3]$ and $z_{\rm s}\in [0.4,0.5]$. The
marginalization intervals are chosen to be consistent with realistic
constraints coming from other experiments, especially from the Cosmic
Microwave Background (e.g., Sievers et al. 2002), the 2dF (Lahav et
al. 2001) and the SLOAN (Szalay et al. 2001) results, and photometric
redshifts. Note that we always fix the cosmological constant to
$\Omega_\Lambda=1-\Omega_{\rm m}$, that is we assume the (correct)
flat geometry. The strongest constraints are found for the
$\Omega_{\rm m}$-$\sigma_8$, $\Gamma$-$\sigma_8$ and $\Omega_{\rm
m}$-$\Gamma$ pairs, suggesting that once the redshift distribution of
the sources is known we can obtain stringent constraints on the
cosmological parameters from cosmic shear. The degeneracy directions
can be better studied with a strong prior likelihood analysis as shown
on Fig.\ts\ref{figstrongprior}.  On this plot, we assume that the two
hidden parameters of each constrained pair are known and fixed at
their true value.  Again we see that the best constraints come from
pairs of parameters which exclude the redshift information. It is
remarkable that the degeneracy among these parameters can be broken
quite efficiently, given that the cosmic shear signal has so few
``spectral features'', compared to the cosmic microwave background for
instance. In fact, the degeneracy can be broken when using linear and
non-linear scales simultaneously (as discussed in Jain \& Seljak
1997).  One would obtain an even better degeneracy breaking by going
further into the linear regime (up to $1$ or $2$ degrees). Also, it is
interesting to note that the parameters which are least affected
by the marginalization (compare Figures \ref{figweakprior} and
\ref{figstrongprior}) is the $\Omega_{\rm m}$-$\sigma_8$ pair, which
is most directly comparable to cluster normalization constraints
(see a discussion on the comparison between cluster normalization and
cosmic shear constraints in van Waerbeke et al. 2002).

\section{Conclusions}
In this paper we have obtained general expressions for the covariance
of an estimator for the shear correlation function as it is determined
from cosmic shear data. Using the approximation that the four-point
function of the shear separates in products of two-point function, the
covariance can be expressed directly in terms of the correlation
functions, as given in (\ref{eq:Cpp}--\ref{eq:Cpm}) and can, for a
given data set, be calculated directly. The covariance of the
correlation functions depends on the number of pairs that enter their
estimate, which in turn depends on the solid angle covered by the
survey and the survey geometry (see also Kaiser 1998); in addition, it
depends on the intrinsic galaxy ellipticities and the number density
of galaxies.

Next, considering a survey geometry of a single compact region of
solid angle $A$, we have calculated the ensemble average of the
covariances, using approximations which a valid for separations
$\ll\sqrt{A}$. The ensemble average of the covariances can then be
reduced to integrals which are readily evaluated numerically. The
estimate for the correlation function $\xi_-(\vt)$ decorrelates
quickly, i.e. estimates of $\xi_-$ for two angular scales which differ
by more than a factor $\sim 2$ are essentially decorrelated. On the
other hand, the estimates of $\xi_+$ are correlated over much larger
angular scales. The cross-correlation between $\xi_+(\vt_1)$ and
$\xi_-(\vt_2)$ is significant for $\vt_1\lesssim \vt_2$, which is due
to the properties of the different filters with which these
correlation functions are related to the power spectrum
$P_\kappa(\ell)$. 

Using these ensemble-averaged covariances for the correlation
functions, we have obtained the covariances for other two-point
measures of the cosmic shear, primarily the aperture mass dispersion
and the power spectrum. Of particular interest is the reconstruction
of the power spectrum $P_\kappa(\ell)$ from the correlation functions;
we have constructed a simple estimator for $P_\kappa$ and the band
powers of it in terms of the $\xi$'s and found that the band power can
be obtained with surprisingly large accuracy from even a
moderately-sized cosmic shear survey. Finally, we have investigated
the confidence regions for the most relevant cosmological parameters
($\Omega_{\rm m}$, $\sigma_8$, $\Gamma$ and $z_{\rm s}$) with a maximum
likelihood approach. We studied our ability to constrain
simultaneously these parameters from a measurement of the shear
correlation function, as well as the effect of some level of lack of
knowledge using the marginalization technique.
 
In a future paper, we shall investigate strategies for conducting
cosmic shear surveys by optimizing the survey geometry.

\begin{acknowledgements}
This work was supported by the TMR Network ``Gravitational Lensing:
New Constraints on Cosmology and the Distribution of Dark Matter'' of
the EC under contract No. ERBFMRX-CT97-0172, by the German Ministry
for Science and Education (BMBF) through the DLR under the project 50
OR 0106, and by the Deutsche Forschungsgemeinschaft under the project
SCHN 342/3--1.

\end{acknowledgements}

\end{document}